\documentclass[11pt]{article} 
 
\usepackage{epsfig}  
\usepackage{amsmath}
\usepackage{amssymb}

\newcommand{\D}{\displaystyle}

\begin{document} 

\begin{titlepage}

\vspace{2cm}

\title
{\vskip -70pt
\begin{flushright}
{\normalsize \ DAMTP--2000--120\\[-2mm]{\tt arXiv:hep-th/0010277}}\\
\end{flushright}
\vskip 2cm
{\bf Quantum Chern--Simons vortices on a sphere}
\vspace{1.5cm}
}

\author{
{\Large \sc N. M. Rom\~ao}
\thanks{e-mail: {\tt N.M.Romao@damtp.cam.ac.uk}}\\[7mm]
{\normalsize \sl Department of Applied Mathematics and Theoretical Physics}\\
{\normalsize \sl Centre for Mathematical Sciences, University of Cambridge} \\
{\normalsize \sl Wilberforce Road, Cambridge CB3 0WA, England}\\
\\}

\date{October 27, 2000}

\maketitle
\thispagestyle{empty}
\vspace{1cm}

\begin{abstract}

\noindent
The quantisation of the reduced first-order dynamics of the 
nonrelativistic model 
for Chern--Simons vortices introduced by Manton is studied on a sphere
of given radius.
We perform geometric quantisation on the moduli space of static solutions,
using a K\"ahler polarisation, to construct the quantum 
Hilbert space. Its dimension
is related to the volume of the moduli space in the usual classical 
limit.
The angular momenta associated with the rotational $SO(3)$ symmetry of 
the model are determined for both the classical and the quantum
systems. The results obtained are consistent with the interpretation 
of the solitons in the model as interacting bosonic particles.

\end{abstract}

\end{titlepage}

\section{Introduction}

Over the past decade, much attention has been given to 
($1+2$)-dimensional field theories including a Chern--Simons term. 
The pure Chern--Simons gauge theory, although still interesting 
both from the mathematical and the physical points of view, has no 
dynamics by itself. However, many interesting models for field dynamics 
can be obtained by adding to the Chern--Simons action 
Maxwell or Yang--Mills terms and/or 
interactions with other fields \cite{Du}. Some of these models have
been shown to admit classical solitonic solutions (vortices), at least
for critical or ``self-dual'' values of the parameters in the lagrangian.
These objects can be regarded as smeared-out particles which retain a
characteristic size and superpose nonlinearly;
unlike some types of solitons, they can also be assigned a pointlike
core individually. 
In specific models, vortices often turn out to possess
rather exotic properties, which may be relevant in applications.
For example, models with abelian vortices have been important in
attempts to explain phenomena in condensed matter theory such as 
superconductivity and the fractional quantum Hall effect.

In \cite{Mfovd}, Manton constructed a nonrelativistic lagrangian
for a $U(1)$ gauge field minimally coupled to a complex scalar on the
plane which 
describes vortex dynamics. The action for the gauge field 
includes a Chern--Simons term and the purely spatial part of the
Maxwell action. The equations of motion for the field theory 
are first-order in time, and 
they admit the well-known Bogomol'ny\u \i\ vortices \cite{JT} of the 
Ginzburg--Landau theory as static solutions, for special values of the
parameters. Mathematically, Bogomol'ny\u\i\ vortices are rather well 
understood, even though on general surfaces they cannot be constructed
analytically. 
Their space of gauge equivalence classes splits into disjoint 
sectors ${\cal M}_{N}$
labelled by an integer vortex number $N\in \mathbb{Z}$, each ${\cal M}_{N}$
(the moduli space of $N$ vortices) being a smooth $2|N|$-dimensional manifold.
In his paper, Manton explored the dynamics of time-dependent fields by 
explicitly reducing the field theory lagrangian to an effective 
(finite-dimensional) mechanical system on the moduli space ${\cal M}_{N}$.
The lagrangian equations of motion for the reduced system
are again first-order in time, 
and so the moduli space is to be regarded as the phase space where a 
noncanonical hamiltonian dynamics takes
place. The symplectic form defining the dynamics, determined by the
kinetic term, contains nontrivial 
information about the system; of course, it may still be written in 
canonical form locally, but not in a natural way.
Time evolution is determined by the potential energy alone, which is
supposed to be small so that the field configurations are still
approximately Bogomol'ny\u\i\ vortices.
In the case of two vortices, the reduced system describes namely a
rigid uniform 
rotation of the two vortex cores about their midpoint, with an angular 
velocity which is maximal when the distance between the two is roughly
a vortex diameter.

If we place the vortices on a compact surface $\Sigma$, with a
riemannian metric and an orientation, rather than on the
plane, the moduli spaces ${\cal M}_{N}$ also become compact. 
The metric on $\Sigma$ fixes a complex structure, which in turn
induces a complex structure on ${\cal M}_{N}$.
This complex structure can be shown to be compatible with 
the symplectic form relevant for the dynamics, and so each ${\cal M}_{N}$
becomes a K\"ahler manifold.
Compact K\"ahler phase spaces are optimal stages for geometric 
quantisation (cf. \cite{Sni}, \cite{Woo}). 
The complex geometry supplies a natural (K\"ahler) polarisation, for
which the corresponding quantum Hilbert space turns out to be 
finite-dimensional.
This approach to the quantisation of the vortex system is to be
included in a more general framework, pioneered 
in \cite{GM} in the context of BPS monopoles. 
The idea is to probe the quantum behaviour of solitons through
geometric quantisation of the reduced dynamics on the moduli space of
static solutions, when such a space is available.
In the more familiar situation of the abelian Higgs model \cite{Sam},
where the reduced system is of
second order, there is a canonical hamiltonian description of the
classical dynamics and the quantisation can be carried out using the
vertical polarisation of $T^{*}{\cal M}_{N}$, which leads to a truncated
Schr\"odinger representation of the quantum system. The accuracy of
the approximation involved is very difficult to assess, and the study
of an example where the Schr\"odinger representation is not available,
as is the case here, is of
considerable interest. From the point of view of geometric
quantisation itself, it is fortunate that Manton's system
seems to provide us with a nontrivial example where it may be put to 
work rather directly.

The main aim of the present work is to discuss the
geometric quantisation of Manton's reduced system of periodic vortex 
dynamics, when space is taken to be a sphere of a given radius~$R$. In
particular, we shall determine the dimension of the Hilbert space
and construct the quantum operators corresponding to the 
conserved angular momenta determined by the natural action of $SO(3)$.

Let us summarise how this paper is organised.
In section 2, we describe the generalisation of Manton's model
to the case where the spatial surface is compact. 
In section 3, we gather some results concerning the moduli space of
Bogomol'ny\u \i\ vortices on a sphere and its use in the study of
the field theory dynamics. 
The effective lagrangian on the moduli space is constructed in section 4.
In section 5, we obtain conservation laws for both the field 
theory and the reduced dynamical system, and show that they are
consistent.
In section 6, the setup for the geometric quantisation
of the reduced dynamics on the moduli space of static 
solutions is presented. 
The dimension of the Hilbert space of wavefunctions is computed in 
section 7, and we show how to construct the quantum angular momentum 
operators in section 8.
Finally, we discuss the results obtained and address some outstanding
issues.

\section{First-order Chern--Simons vortices}

We start by discussing the generalisation of the model introduced by
Manton in \cite{Mfovd} to the situation where space is compact. We
shall consider time-periodic boundary conditions in the formulation of
the variational principle, and accordingly we fix space-time to be of 
the form $S^{1} \times \Sigma$, where $\Sigma$ is a compact and oriented
2-dimensional riemannian manifold. 
In the remaining sections of this paper we will assume that $\Sigma$ 
is a sphere, but for now this restriction is unnecessary.
Since in two dimensions any metric is conformally
flat, we can introduce a complex coordinate $z$ locally on $\Sigma$
and write
\begin{equation}\label{metric}
ds^{2} = dt^{2}-\Omega^{2}(z,\bar{z})dzd\bar{z}
\end{equation}
where $t$ parametrises time.

Naively, the lagrangian we would like to consider is
\begin{eqnarray} \label{lagr}
{\cal L} [A,\phi] & = & \gamma
\left(\frac{i}{2}(\bar{\phi}D_{t}\phi-\phi\overline{D_{t}\phi})
-A_{t}\right)\Omega^{2}
+\mu\left( B A_{t}+2i(E_{z}A_{\bar{z}}-E_{\bar{z}}A_{z})  \right)\\ \nonumber
& & -\left(\frac{1}{2} B^{2}\Omega^{-2}
+ \left(|D_{z}\phi|^{2}+ |D_{\bar{z}}\phi|^{2}\right)
+\frac{\lambda}{8}\left( 1-|\phi|^{2}\right) ^{2}\Omega^{2}\right).
\end{eqnarray}
This reduces to Manton's lagrangian \cite{Mfovd} without transport current
if we take $\Sigma$ to be the plane and set $\Omega^{2}=1$. 
Here, $A=A_{t}dt+A_{z}dz+A_{\bar{z}}d\bar{z}$ is the real-valued
$U(1)$ gauge field, with curvature 
$F_{A}=dA=(E_{z}dz+E_{\bar{z}}d\bar{z}) \wedge 
dt+\frac{i}{2}Bdz\wedge d\bar{z}$, and $\phi$ the Higgs scalar field.
The covariant derivatives are 
$D_{\nu}\phi=\partial_{\nu}\phi-iA_{\nu}\phi$, where
$\nu=t$, $z$ or $\bar{z}$.
Of course, $\frac{i}{2}{\cal L}$ is to be thought of as the coefficient of a
3-form on some open subset of $S^{1} \times \Sigma$ where local expressions 
for the fields can be given, and the factors of $\Omega$ introduced in
(\ref{lagr}) are imposed by the natural interpretation given to the 
different terms.

To set up the classical field theory, we must give as global data a
principal $U(1)$ bundle ${P}$ over $S^{1}\times\Sigma$. With respect
to local trivialisations, the gauge
field is interpreted as
a connection on this bundle, the Higgs field as a section of 
the complex line bundle associated to ${P}$ by the defining 
representation, and $\bar{\phi}$ is a section of the bundle dual to this one.
It is natural to restrict to the situation 
where ${P}$ is the pull-back to $S^{1} \times \Sigma$ of a 
$U(1)$ bundle on $\Sigma$; in particular, the transition functions
will be time-independent. Topologically, $U(1)$ bundles on a compact
surface are classified by their first Chern class $N\in \mathbb{Z}$, 
which can be interpreted as the net number of units of quantised magnetic
flux through space at any time,
\begin{equation}\label{qflux}
\frac{i}{2} \oint_{\Sigma}Bdz\wedge d\bar{z}=2\pi N.
\end{equation}
We may assume without loss of generality that the bundle we are
considering over $\Sigma$ can be trivialised on an open disc 
$U_{1}\subset \Sigma$ and on an open neighbourhood $U_{2}$ of its
complement, with $U_{1}\cap U_{2}$ being a very narrow annulus
which for most purposes can be identified with its retraction
$\partial U_{1}$. More precisely, we may have to 
consider sub-patches of $U_{2}$ to make sense of local data such as
the relevant coordinate $z$, but this will not affect the discussion
of the aspects related to the nontriviality of ${P}$ which will be 
our main concern, because ${P}|_{U_{2}}$ is trivial. Thus we shall 
consider ${P}$ to be defined by the homotopy class of a single
transition function $f_{12}:\partial U_{1} \rightarrow U(1)$ whose 
degree is $N$, and we refrain from introducing partitions of unity 
to keep the discussion as simple as possible.

The term with coefficient $\mu$ is the Chern--Simons
density $\mu A\wedge dA$. On the overlap of the two trivialising
patches $U_{1}$, $U_{2}$,
\[
A^{(1)}\wedge dA^{(1)} =
A^{(2)}\wedge dA^{(2)} -if_{12}^{-1}df_{12}\wedge dA^{(2)}
\]
where $A^{(j)}$ denotes the connection 1-form on $U_{j}$,
and so its values on the trivialising patches do not agree on the
overlap $U_{1}\cap U_{2}$. 
So in general we cannot define an action by just using partitions of 
unity to patch together pieces of the lagrangian given by
(\ref{lagr}), as we can do for gauge-invariant lagrangians. 
Notice that the term proportional to $A_{t}$, although gauge-dependent, 
is unambiguously defined globally since we are assuming that the transition 
functions are time-independent.
The most elegant way to define the Chern--Simons action is as the 
integral of the gauge-invariant second Chern form $\mu dA\wedge dA$, on any
4-dimensional manifold $M$ with boundary
$\partial {M} = S^{1}\times \Sigma$. 
Here, $A$ is a connection on a principal $U(1)$ bundle on ${M}$
which restricts to our bundle $P$ on $S^{1}\times \Sigma$.
There is no obstruction to the existence of such an extension
of $P\rightarrow S^{1}\times \Sigma$, since in our case it
would lie in the group $H_{3}(\mathbb{CP}^{\infty};\mathbb{Z})$,
which is trivial (cf.~\cite{DW}). The action should be independent 
(mod $2\pi$) of the choice of the manifold ${M}$ and the bundle over it, 
and this imposes the constraint
\begin{equation} \label{mu}
\mu\in \frac{1}{4 \pi} \mathbb{Z}
\end{equation}
on the Chern--Simons coefficient. We shall write $\kappa:=4\pi \mu$.

The group of gauge transformations ${\cal G}$ consists of smooth maps 
from $S^{1}\times \Sigma$ to $U(1)$. The connected component of the 
identity ${\cal G}^{0}$ is the subgroup of maps homotopic to the
identity (the small gauge transformations), and the connected
components of ${\cal G}$ are labelled
by 2-homology classes of space-time, dual to the 1-cycles around which
the gauge transformations have nontrivial winding:
\begin{equation}\label{ccompts}
{\cal G}/{\cal G}^{0}\simeq H_{2}(S^{1}\times \Sigma; \mathbb{Z})
\simeq \mathbb{Z} \oplus \mathbb{Z}^{\oplus 2 {\tt g}}
\end{equation}
Here, ${\tt g}$ is the genus of $\Sigma$, and we can choose for the 
generator ${\sigma}$ of the first $\mathbb{Z}$ factor the class of a
positively-oriented copy of $\Sigma$ at a particular instant. It can
be shown \cite{DGS}
that a gauge transformation in the connected component of ${\cal G}$
labelled by a class whose first component in the above decomposition
is $k\sigma$ has the effect of adding the term $2\pi\kappa k N$ to
the Chern--Simons action, so that this action is gauge-invariant 
(mod $2\pi$) if and only if the condition (\ref{mu}) holds.

It is possible to express the Chern--Simons action entirely in terms
of the 3-dimensional data by treating carefully the boundary terms
of the 4-dimensional Chern action introduced above, as shown in \cite{DGS}. 
The result is that we should add a correction to the sum of the 
integrals of the Chern--Simons bulk term appearing in (\ref{lagr})
over $U_{1}$ and $U_{2}$. The correction term is the double integral
\begin{equation} \label{correct}
\mu i \oint_{S^{1}\times \partial U_{1}}f_{12}^{-1}df_{12}\wedge A^{(1)}.
\end{equation}

We still have to ensure that the term proportional to $A_{t}$ in the
action is gauge-invariant (${\rm mod}\; 2\pi$). Under a gauge
transformation $g$, $A_{t}$ changes as
\[
A_{t} \mapsto
A_{t} -i g^{-1}\partial_{t}g.
\]
If the class of $g$ in the first factor of ${\cal G}/{\cal G}^{0}$
in (\ref{ccompts}) is $k\sigma$, then everywhere on $\Sigma$
\[
i\oint_{S^{1}} g^{-1} \partial_{t}g dt=k 
\]
and the change in the action is
\[
-\frac{i}{2}
\gamma k \oint_{\Sigma}\Omega^{2}dz\wedge d\bar{z}=-\gamma k\, 
{\rm Vol}(\Sigma).
\]
This will be in $2\pi\mathbb{Z}$
for all $k\in \mathbb{Z}$ if and only if we impose the constraint
\begin{equation}\label{gamma}
\gamma \,{\rm Vol}(\Sigma) \in \mathbb{Z}.
\end{equation}

The action for Manton's model on $\Sigma$ can then be written as
\begin{equation} \label{action}
S[A,\phi]=\sum_{j=1}^{2} \int_{S^{1}\times U_{j}} 
{\cal L}[A^{(j)}, \phi^{(j)}] d^{3}x
+\mu i\oint_{S^{1}\times \partial U_{1}}f_{12}^{-1}df_{12}\wedge A^{(1)}
\end{equation}
where we impose the constraints (\ref{mu}) and (\ref{gamma}) to the
classical parameters to ensure that $e^{iS}$ is well defined and 
gauge invariant. 
To implement the variational principle, we consider variations $\delta
A$, $\delta \phi$ and $\delta \bar{\phi}$ of the fields which are 
a 1-form and sections of the bundles 
associated to ${P}$ by the fundamental representation and its dual,
respectively. 
As usual, the variation of the first (bulk) term in (\ref{action}) 
yields after integration by parts
\begin{equation}\label{parts}
\delta S = \sum_{\psi}\sum_{j=1}^{2}\left\{
\int_{S^{1}\times U_{j}} \left[
\frac{\delta{\cal L}}{\delta \psi}-\partial_{\nu}
\left(\frac{\delta {\cal L}}{\delta\partial_{\nu}\psi}\right)
\right]\delta \psi \,d^{3}x
+
\int_{S^{1}\times U_{j}}
\partial_{\nu}\left(\frac{\delta{\cal L}}
{\delta \partial_{\nu}\psi}\delta \psi \right)d^{3}x
\right\}
\end{equation}
Here, $\psi$ is any of $A_{\nu}$, $\phi$ or $\bar{\phi}$, and the
($j$) subscripts have been suppressed. If we define on each $U_{j}$ the 
1-form
\[
\Psi:=\Omega^{-2}\left( \frac{\delta {\cal L}}{\delta 
\partial_{t} \psi}\right) dt-
\left( \frac{\delta {\cal L}}{\delta 
\partial_{\bar{z}} \psi}\right)dz-
\left( \frac{\delta {\cal L}}{\delta 
\partial_{z} \psi}\right) d\bar{z}
\]
and denote the Hodge star of (\ref{metric}) by $*$, then the last
integral in the expression (\ref{parts}) can be written as
\begin{equation}\label{bterm}
\int_{S^{1}\times U_{j}}\partial_{\nu}\left(\frac{\delta{\cal L}}
{\delta \partial_{\nu}\psi}\delta \psi \right)d^{3}x=
\oint_{S^{1}\times \partial U_{j}}
*\Psi\,\delta \psi.
\end{equation}
It is easy to verify by direct computation that, for the terms in
${\cal L}$ which are locally gauge-invariant, the contributions to the
components of the 2-form $*\Psi\delta\psi$ are just functions on each
trivialising patch. Therefore, their corresponding $j=1,2$
contributions in (\ref{bterm}) cancel as they should, since the boundaries 
$\partial U_{1}$ and
$\partial U_{2}$ are very close but carry opposite orientations. For 
the term proportional to $A_{t}$, the values of the contributions on 
the $j=1,2$ patches also cancel because of our assumption of the
time-independence of $f_{12}$. However, the Chern--Simons term has
a nonvanishing contribution. Indeed,
\[
\delta(A\wedge dA)=2dA\wedge \delta A-d(A\wedge \delta A)
\] 
and so the sum of the $j=1,2$ contributions to (\ref{bterm}) is
\[
\mu\oint_{S^{1}\times \partial U_{1}}
(A^{(1)}\wedge\delta A-A^{(2)}\wedge \delta A)=
-\mu i \oint_{S^{1}\times \partial U_{1}}f_{12}^{-1}df_{12}
\wedge\delta A,
\]
which exactly cancels the variation of the correction (\ref{correct})
to the Chern--Simons bulk term. So we conclude that the stationarity
of the action (\ref{action}) is exactly expressed by the
Euler--Lagrange equations for the naive lagrangian (\ref{lagr})
in each trivialising coordinate patch. They read
\begin{eqnarray}
&\gamma i D_{0}\phi=-(D_{z}D_{\bar{z}}\phi+D_{\bar{z}}D_{z}\phi)\Omega^{-2}-
\frac{\lambda}{4}\left(1-|\phi|^{2} \right) \phi & \label{nlSchr}\\
&\partial_{z}(B\Omega^{-2})=-iJ_{z}+2\mu E_{z} & \label{gAmp}\\
& 2\mu B=\gamma(1-|\phi|^{2})\Omega^{2} & \label{GauB}
\end{eqnarray}
where the gauge-invariant supercurrent $J_{z}$ is defined by
\begin{equation} \label{superc}
J_{z}:=-\frac{i}{2}\left(\bar{\phi}D_{z}\phi-
\phi \overline{D_{\bar{z}}\phi}\right).
\end{equation}
We are making the assumption that these equations admit nonstatic 
time-periodic solutions which can be patched together on $\Sigma$. 
Notice that no higher than first-order time derivatives of the fields 
appear in (\ref{nlSchr})--(\ref{GauB}).
Equation (\ref{nlSchr}) is a gauge-invariant nonlinear Schr\"odinger 
equation for $\phi$, while (\ref{gAmp}) is a version of Amp\`ere's law and 
(\ref{GauB}) can be interpreted as a magnetic Gauss' law.

\section{Static vortices on a sphere}

Static configurations are time-independent solutions of the field 
equations of motion with vanishing $A_{t}$. For them, our lagrangian 
reduces to the
Ginzburg--Landau energy functional ${\cal E}$, given by the last 
bracket in~(\ref{lagr}). Notice that this is also the functional
relevant to the discussion of static solutions in the abelian Higgs
model.
If $\lambda=1$, one can show that all the critical points of ${\cal E}$ 
satisfy the two-dimensional Bogomol'ny\u \i\ equations \cite{Bo} on 
$\Sigma$. For configurations with $N>0$, these read
\begin{eqnarray}
&D_{\bar{z}}\phi=0 &\label{covhlm} \\
&2 B=(1-|\phi|^{2})\Omega^{2}&\label{Bogom}
\end{eqnarray}
and their solutions are known as vortices. If $N<0$, they are called
antivortices and satisfy similar equations, with $D_{\bar{z}}$ 
replaced by $D_{z}$ in (\ref{covhlm}), whereas a minus sign is introduced
in (\ref{Bogom}). 
Solutions $(A,\phi)$ to the Bogomol'ny\u\i\ equations for which (\ref{qflux})
holds have 
energy ${\cal E}=|N|\pi$. 
Henceforth, we shall be interested in the $N>0$ case only.
Write $\phi=e^{\frac{1}{2}h+i\chi}$; the function $h$ is gauge-invariant
while $\chi$ (defined only modulo $2\pi$) is not, and both are real.
Equation (\ref{covhlm}) can be used to obtain $A_{z}$ in terms of $h$
and $\chi$, and substitution in (\ref{Bogom}) yields
\begin{equation} \label{newBogom}
4\partial_{z}\partial_{\bar{z}}h-(e^{h}-1)\Omega^{2}=
4\pi\sum_{r=1}^{N}\delta^{(2)}(z-z_{r}).
\end{equation}
The solution to this equation provides all the information needed to 
reconstruct the fields on $\Sigma$; $\chi$ has to give
the required winding properties of $\phi$ in each patch, but the gauge
freedom leaves it otherwise undetermined.

The Bogomol'ny\u \i\ equations were first studied on a compact Riemann 
surface $\Sigma$ by Bradlow \cite{Br} (cf.~also \cite{GP} and
references therein). 
He showed existence and  uniqueness of a 
solution satisfying (\ref{qflux}) with the Higgs field having exactly 
$N$ zeros (counted with well-defined
multiplicities) in any configuration, provided
\begin{equation}\label{Brad}
{\rm Vol}(\Sigma) > 4\pi N.
\end{equation}
Thus, given the topological constraint (\ref{qflux}), and
if the bound (\ref{Brad}) is satisfied, the moduli space of
solutions to the Bogomol'ny\u\i\ equations 
can be described as the
symmetric product ${\cal M}_{N}=S^{N}\Sigma=\Sigma^{N}/\mathfrak{S}_{N}$, 
a smooth $2N$-manifold. These solutions can be interpreted as
nonlinear superpositions of $N$ indistinguishable vortices
located at the zeros of the Higgs field (the vortex cores), which play
the r\^ole of moduli.

For the rest of this paper, we shall restrict to the situation where
$\Sigma$ is a 2-sphere of radius~$R$, which for later convenience we
assume to be centred at the origin of $\mathbb{R}^{3}$. We choose
the open subsets $U_{1}$ and $U_{2}$ introduced in section~2 to be
discs around the North and South poles
respectively, where stereographic coordinates can be used.
In the next section, we will argue that actually we can shrink $U_{2}$
to a point, and focus entirely on a $U_{1}$ which covers all of $\Sigma$
except the South pole. So we can parametrise the position of the
vortices in the open dense $U_{1}$ by a coordinate $z$ with inverse 
\begin{equation} \label{cartes}
z\mapsto R\left(   
\frac{z+\bar{z}}{1+|z|^{2}},
-i\frac{z-\bar{z}}{1+|z|^{2}},
\frac{1-|z|^{2}}{1+|z|^{2}}
\right).
\end{equation}
As usual, we write $z=\infty$ to refer to the South pole.
In terms of this coordinate, the conformal factor in (\ref{metric}) is just
\begin{equation}\label{conffactor}
\Omega^{2}(z,\bar{z})=\frac{4R^{2}}{\left(1+|z|^{2}\right)^{2}}.
\end{equation}

The positions $z_{1},\ldots, z_{N}$ of $N$ vortices define coordinates
almost everywhere on the moduli space. They are regular only
in the subset of configurations for which all the zeros of $\phi$ 
are simple and none of them occur at the South pole. To 
parametrise the whole subset 
$V_{0}\subset{\cal M}_{N}$ of static configurations with no zeros of
the Higgs field at $z=\infty$, we can introduce instead the 
elementary symmetric polynomials in the $N$ variables $z_{r}$:
\begin{equation}\label{esses}
s_{k}:=s_{k}^{[N]}(z_{1},\ldots,z_{N})
=\sum_{1\le j_{1}<\cdots<j_{k}\le N}z_{j_{1}}\cdots z_{j_{k}}
,\;\;\;\;\;\;\;1\le k\le N.
\end{equation}
More generally, let $V_{j}\subset{\cal M}_{N}$ denote the subset of
configurations with exactly $j$ vortices at $z=\infty$; it is
parametrised by the symmetric polynomials $s_{k}^{[N-j]}$ of the 
coordinates of the $N-j$ remaining vortices. Clearly, 
${\cal M}_{N}=\coprod_{j=0}^{N}V_{j}$
gives a decomposition of ${\cal M}_{N}$ into $N+1$ disjoint $2(N-j)$-cells
$V_{j}\cong\mathbb{C}^{N-j}$, and it is easy to verify that they are
glued together so as to give ${\cal M}_{N}=\mathbb{CP}^{N}$.

Later on, it will be useful to consider the standard decomposition
of the moduli space as union of affine pieces 
${\cal M}_{N}=\cup_{j=0}^{N}W_{j}$, where the $W_{j}$ are defined
in terms of homogeneous coordinates as usual,
\[
W_{j}:=\{(Y_{0}:Y_{1}:\cdots:Y_{N})\,|\,Y_{j}\ne 0 \}\cong \mathbb{C}^{N}.
\]
We may set $W_{0}=V_{0}$ say, and identify the $s_{k}$ in
(\ref{esses}) with the inhomogeneous coordinates of $W_{0}$,
\[
s_{k}=y^{(0)}_{k}:=\frac{Y_{k}}{Y_{0}},\;\;\;\;\;\;\;1\le k\le N.
\]
Letting $s_{0}:=1$, we can relate the inhomogeneous coordinates in the
other $W_{j}$ to the $s_{k}$ by
\[
y_{k}^{(j)}:=\frac{Y_{k}}{Y_{j}}=\frac{s_{k}}{s_{j}},
\;\;\;\;\;\;0\le k \le N,\;k\ne j.
\]
It is not hard to see that $V_{j}$ is being identified with the $(N-j)$-plane
$y^{(j)}_{0}=\ldots=y^{(j)}_{j-1}=0$ in~$W_{j}$.

We still need to introduce a further piece of notation. 
Near the position $z=z_{r}$ of an isolated vortex, 
$h={\rm log}|\phi|^{2}$ can be expanded as
\cite{Str}, \cite{Sam}
\begin{equation}\label{bs}
h(z;z_{1},\ldots,z_{N})=
\log|z-z_{r}|^{2} + a_{r} + \frac{b_{r}}{2}(z-z_{r}) + \frac{\bar{b}_{r}}{2}
(\bar{z}-\bar{z}_{r})+O(|z-z_{r}|^{2}).  
\end{equation}
Here, $a_{r}$ and $b_{r}$ are functions of the positions
$z_{1},\ldots, z_{N}$ of all the $N$ vortices.
If $N=1$, spherical symmetry can be used to show that \cite{Msmv}
\[
b_{1}=-\frac{2\bar{z}_{1}}{1+|z_{1}|^{2}}.
\]
This function describes how the level curves of $|\phi|^{2}$, 
which consist of
circles centred around the core of the vortex on the sphere, are 
distorted by the stereographic projection onto the $z$-plane. 
An analogous situation arises when we consider more generally the 
subvariety ${\cal M}_{N}^{\rm co}\subset{\cal M}_{N}$ of 
configurations of $N$ coincident
vortices. It is parametrised by the position $z=Z$ of the only zero 
of the Higgs
field, whose modulus squared has a logarithm with an 
expansion identical to (\ref{bs})
around $Z$, but with a coefficient $N$ before the logarithmic term; the
coefficient of $\frac{1}{2}(z-Z)$ is then
\begin{equation}\label{bN}
b=-\frac{2N\bar{Z}}{1+|Z|^{2}}.
\end{equation}
When the positions of the $N$ vortices do not coincide, there is an 
additional distortion
to the $|\phi|^{2}$ contours caused by the mutual interactions, and 
this leads to nontrivial $b_{r}$ coefficients in (\ref{bs}).
It turns out that the functions $b_{r}$ contain all the information about 
the interactions relevant for the kinetic term of the reduced
mechanical system. This was also the case in Samols' analysis of
the abelian Higgs model~\cite{Sam}.

It will be useful later to relate the functions $b_{r}$ to their
pullbacks
\[
T^{*}b_{r}(z_{1},\ldots,z_{N}):=b_{r}(T(z_{1}),\ldots,T(z_{N})) 
\]
under an isometry $T$ of the sphere. This relationship is
easy to obtain directly from the expansion (\ref{bs}) by requiring
that $h$ be invariant:
\begin{equation}\label{btransf}
T^{*}b_{r}= \frac{b_{r}}{T'(z_{r})}-
\frac{T''(z_{r})}{\left(T'(z_{r})\right)^{2}}.
\end{equation}
If we interpret $T$ as a (passive) change of coordinates, this equation 
can be regarded as the formula for the transformation of Christoffel 
symbols for an affine connection, as pointed out in \cite{MNvvms}. 
In fact, we shall define a natural unitary 
connection on a line bundle over the moduli space of static 
vortices. In the next section, we show that this connection is the 
essential ingredient of the reduced lagrangian for Manton's model.
The functions $b_{r}$ are part of the coefficients of the connection
1-form, and they incorporate the interactions among the vortices 
at zero potential energy into the reduced dynamics.

\section{The effective mechanical system}

Ultimately, we are interested in understanding the dynamics of the
field configurations, and this is a much harder
problem than the study of static solutions. However, it is natural
to expect that, for $\lambda$ close to 1, 
slow-varying solutions of the field theory 
should be well approximated by static solutions evolving along the
directions defined by the linearised Bogomol'ny\u\i\ equations.
This idea was introduced by Manton in the context of
Yang--Mills--Higgs monopoles
\cite{MBPS} and has proven very fruitful in many situations. 
For the abelian Higgs model, 
a careful analysis by Stuart \cite{StuAH} showed that this so-called
adiabatic approximation is exponentially
accurate in the limit of small velocities, and that it holds even when
the coupling differs slightly from the self-dual value $\lambda=1$. 
The field dynamics of $N$ vortices is reduced to an effective 
mechanical system on the moduli space ${\cal M}_{N}$. 
The relevant lagrangian can be obtained by evaluating the 
lagrangian for the field theory at static solutions 
with time-dependent moduli and then integrating out the space dependence.
Samols showed in \cite{Sam} that the reduced dynamics at self-dual coupling 
corresponds to geodesic motion on the moduli space, with respect to a 
K\"ahler metric that encodes information about the local behaviour of the
Higgs field at its zeros; we shall explain this more precisely at the
end of this section.
For general values of $\lambda\simeq 1$, the geodesic motion is
distorted by conservative forces, which are absent at self-dual coupling.

In what follows, we shall study Manton's model within the adiabatic 
approximation, in
the regime $\lambda\simeq 1$ and $\gamma=\mu$. The latter assumption
enables one to obtain a neat expression for the reduced kinetic energy
term, as will be shown below. Notice that, when 
$\gamma=\mu$, equation (\ref{GauB}) reduces to (\ref{Bogom}), and
this raises our hope that the adiabatic picture is a good
approximation to the field theory in this model, as Manton pointed
out in \cite{Mfovd}. Another check is provided by the existence of
consistent conservation laws, which we shall explore in section~5.

We shall follow the analysis in \cite{Mfovd} to obtain the
Lagrangian for the reduced mechanical system on the moduli space.
When $\gamma$ and $\mu$ are set to be equal,
consistency of the conditions (\ref{mu}) and (\ref{gamma}) is 
expressed by
\begin{equation} \label{consist}
\frac{\kappa}{4\pi} {\rm Vol}(\Sigma)=\kappa R^{2} \in  \mathbb{Z}.
\end{equation}
The kinetic energy consists of the terms in (\ref{action}) which
contain time derivatives and $A_{t}$. After using (\ref{Bogom}), it
can be written in the form
\[
T=\frac{i\gamma}{2} \sum_{j=1}^{2}
\int_{U_{j}} {\rm Im} \left( 4 A_{\bar{z}}\dot{A}_{z}-
\bar{\phi} \dot{\phi} \Omega^{2} \right) dz\wedge d\bar{z},
\]
where the overdots denote time derivatives and $z$ is the relevant
stereographic coordinate on each of the discs $U_{j}$. 
Notice that the correction to the naive 
Chern--Simons action has now been cancelled by a boundary term coming 
from the bulk.
In terms of the functions $h$ and $\chi$ introduced in section~3,
we can write
\begin{equation}\label{complT}
T=\frac{i\gamma}{2}\sum_{j=1}^{2}\int_{U_{j}}\left(
2i(\partial_{\bar{z}}\zeta_{z}-\partial_{z}\zeta_{\bar{z}})
+\partial_{t}
(\partial_{z} h \partial_{\bar{z}} \chi + 
\partial_{\bar{z}}h \partial_{z}\chi)
-\dot{\chi}\Omega^{2}
\right) dz\wedge d\bar{z}
\end{equation}
where $\zeta_{z}$, $\zeta_{\bar{z}}$ are the components of the 1-form
\[
\zeta=\dot{\chi}(d\chi + \star dh)+
\frac{1}{4}\dot{h}\,dh
\]
in each trivialising coordinate patch; here, $\star$ is the Hodge star
of the metric on $\Sigma$.
To evaluate (\ref{complT}), we cut discs of small radius $\epsilon$ 
around each vortex position (where the integrand is singular) and
apply Stokes' theorem, letting $\epsilon\rightarrow 0$ at the end.

The only contributions to the integrals around $\partial U_{j}$ come
from the first term in (\ref{complT}), yielding
\[
-i\gamma\oint_{\partial{U_{2}}}\dot{\chi}f_{12}^{-1}df_{12},
\]
where now $\chi$ denotes the argument of 
$\phi^{(2)}$. If we choose $U_{2}$ small
enough so that it does not overlap with any of the trajectories of the
vortices, $\chi$ is globally defined on $U_{2}$. So this term
is a total time derivative and can be discarded from the kinetic
energy.
No other contribution coming from the nontriviality of the bundle ${P}$
arises, and hence we may safely shrink $U_{2}$ to the South pole, while 
$U_{1}\cong\mathbb{C}$ becomes dense in $\Sigma$. Henceforth, $z$ shall
always denote the coordinate in (\ref{cartes}).

To describe the contributions coming from the neighbourhood of the 
vortices, we write near vortex $r$ (cf.~\cite{Mfovd})
\begin{equation} \label{psi}
\chi=\theta_{r}+\psi_{r}, 
\end{equation}
where $\theta_{r}$ is the polar angle in the $z$-plane with respect to
$z_{r}$, and $\psi_{r}$ is a function of the position of the 
vortices only.
Of course, equation (\ref{psi}) assumes that the gauge freedom has 
been reduced in the neighbourhood of the vortices.
The analysis in \cite{Mfovd} goes through unchanged to conclude that
the contributions from the first two terms in (\ref{complT}) add up to
\begin{equation}\label{cont12}
\pi\gamma\sum_{r=1}^{N}\left(2\dot{\psi}_{r}+ib_{r}\dot{z}_{r}
-i\bar{b}_{r}\dot{\bar{z}}_{r}
\right).
\end{equation}
It follows from the expansion (\ref{bs}) that the coefficient $b_{r}$ 
has a singularity when vortex $r$ approaches another vortex $s\ne r$,
\begin{equation}\label{btil}
b_{r}(z_{1},\ldots,z_{N})=2\sum_{s\ne r}\frac{1}{z_{r}-z_{s}}+
\tilde{b}_{r}(z_{1},\ldots, z_{N}),
\end{equation}
where $\tilde{b}_{r}$ is a smooth function.
However, a gauge can be chosen for which each $\psi_{r}$ is given by
\[
\psi_{r}(z_{1},\ldots,z_{N})={\rm arg}\,\prod_{s\ne r}^{N}(z_{r}-z_{s})
\;\;\;\;\;\;\;({\rm mod}\; 2\pi)
\]
and then
\[
\sum_{r=1}^{N}\dot{\psi}_{r}=-i
\sum_{r=1}^{N}\sum_{s\ne r}\left( \frac{\dot{z}_{r}}{z_{r}-z_{s}}-
\frac{\dot{\bar{z}}_{r}}{\bar{z}_{r}-\bar{z}_{s}}\right)
\]
exactly cancels the singularity from the $b_{r}$ coefficients.

The last term in (\ref{complT}) yields again a contribution from the
metric on $\Sigma$. Its time integral gives $-2\pi\gamma$ times the 
signed area enclosed by the trajectories of the vortices $z_{r}(t)$. 
In our local coordinate $z$, the area form can be expressed as
\[
\omega_{\Sigma}=2iR^{2}\frac{dz\wedge d\bar{z}}{\left(1+|z|^{2}\right)^{2}}
=iR^{2}d\left(\frac{zd\bar{z}-\bar{z}dz}{1+|z|^{2}}   \right)
=:d\vartheta.
\]
Therefore, we may write
\begin{equation}\label{cont3}
-2\pi\gamma\oint_{S^{1}}\left(\int_{\Sigma}\dot{\chi}\omega_{\Sigma}\right)dt
=-2\pi\gamma \oint_{S^{1}}
\sum_{r=1}^{N}z_{r}^{*}\vartheta=
2i\pi\gamma R^{2}\oint_{S^{1}} \sum_{r=1}^{N}
\frac{\bar{z}_{r}\dot{z}_{r}-z_{r}\dot{\bar{z}}_{r}}{1+|z_{r}|^{2}}dt
\end{equation}
and interpret the integrand in the last expression above as the
relevant contribution to $T$.

Putting (\ref{cont12}) and (\ref{cont3}) together, we conclude that the
kinetic energy term in the effective lagrangian is given by
\begin{equation}\label{redT}
T^{\rm red}=\pi i \gamma \sum_{r=1}^{N} \left[
\left( 2 R^{2}\frac{\bar{z}_{r}}{1+|z_{r}|^{2}}+\tilde{b}_{r}\right)\dot{z}_{r}
-\left(
2 R^{2}\frac{{z}_{r}}{1+|z_{r}|^{2}}+\bar{\tilde{b}}_{r}\right)
\dot{\bar{z}}_{r}
\right].
\end{equation}
Unfortunately, the potential energy is harder to deal with. It can be
written as
\begin{eqnarray*}
\D V^{\rm red}&=& N \pi+ 
\frac{i}{16}(\lambda-1)\int_{\Sigma}
\left( 1-e^{h} \right) ^{2}\Omega^{2}dz\wedge d\bar{z} \\
&=& N \pi \lambda - \frac{\pi}{2}(\lambda-1) R^{2}  + 
\frac{i}{16}(\lambda-1)\int_{\Sigma}e^{2h}\Omega^{2} dz\wedge d\bar{z}.
\end{eqnarray*}
It does not seem possible to simplify the integral involving 
$e^{2h}$ to a local expression in the moduli. Stuart has shown in 
\cite{Sturr} that, for $N=2$, $V^{\rm red}$ can be approximated by a rational
function of the distance between the two vortices, in the limit where 
Bradlow's bound (\ref{Brad}) comes close to the equality.

The reduced lagrangian $L^{\rm red}=T^{\rm red}-V^{\rm red}$ is
first-order in the time derivatives, just as the field theory
lagrangian. So, as mentioned in the Introduction, 
the equations of motion already arise in hamiltonian form and
the moduli space is to be interpreted as a phase space.
The potential energy $V^{\rm red}$ is the only contribution to the
hamiltonian, while the kinetic term determines a symplectic 
potential ${\cal A}$ in our coordinate patch $W_{0}$ for a noncanonical 
symplectic form on ${\cal M}_{N}$.
Indeed, $T^{\rm red}={\cal A}(\frac{d}{dt})$ with $\frac{d}{dt}$ the
vector field determining time evolution, and ${\cal A}$ the real 1-form
\begin{equation}\label{unitconn}
{\cal A} = \pi i \gamma \sum_{r=1}^{N} \left[
\left( 2 R^{2}\frac{\bar{z}_{r}}{1+|z_{r}|^{2}}+\tilde{b}_{r}\right)dz_{r}
-\left(
2 R^{2}\frac{{z}_{r}}{1+|z_{r}|^{2}}+\bar{\tilde{b}}_{r}\right)
d\bar{z}_{r}
\right].
\end{equation}
This 1-form is regular throughout $W_{0}$.
The equations of motion can be cast as
\[
\iota_{\frac{d}{dt}}\omega=-dV^{\rm red},
\]
where the (global) symplectic form $\omega=d{\cal A}$ is given on $W_{0}$ by
\begin{eqnarray*}
-\frac{4i}{\kappa}\,\omega&=&d\sum_{r=1}^{N}\left[
\left(2 R^{2}\frac{\bar{z}_{r}}{1+|z_{r}|^{2}}+\tilde{b}_{r}\right)dz_{r}
-\left(2 R^{2}\frac{{z}_{r}}{1+|z_{r}|^{2}}+\bar{\tilde{b}}_{r}\right)
d\bar{z}_{r}
\right]\\
&=&\sum_{r=1}^{N}\sum_{s\ne r}
\left(\frac{\partial \tilde{b}_{s}}{\partial z_{r}}dz_{r}\wedge dz_{s}
-\frac{\partial \bar{\tilde{b}}_{s}}{\partial \bar{z}_{r}}
d\bar{z}_{r}\wedge d\bar{z}_{s}\right)-\\
&&-
\sum_{r,s=1}^{N}\left(\frac{\partial b_{r}}{\partial \bar{z}_{s}}+
\frac{\partial \bar{b}_{s}}{\partial z_{r}}+
\frac{4 R^{2} \delta_{rs}}{\left( 1+|z_{r}|^{2}\right)^{2}}\right)
dz_{r}\wedge d\bar{z}_{s}.
\end{eqnarray*}
Notice that we are allowed to replace $b_{r}$ by $\tilde{b}_{r}$
in expressions like $\frac{\partial b_{r}}{\partial \bar{z}_{s}}$,
since it is know from (\ref{btil}) that $b_{r}-\tilde{b}_{r}$ is holomorphic.
Either of the two arguments given in \cite{Sam} shows that locally
\begin{equation}\label{bpotential}
\tilde{b}_{r}=\frac{\partial {\cal B}}{\partial z_{r}}
\end{equation}
for a real function ${\cal B}$. Hence
\begin{equation}\label{omega}
\omega=-\frac{i\kappa}{2}\sum_{r,s=1}^{N}
\left(\frac{2 R^{2}\delta_{rs}}{\left(1+|z_{r}|^{2}\right)^{2}}+
\frac{\partial \bar{b}_{s}}{\partial z_{r}}\right)
dz_{r}\wedge d\bar{z}_{s}.
\end{equation}

It is easily verified that equation (\ref{bpotential}) is equivalent to  
$\omega$ being a K\"ahler form on $W_{0}$ with respect to the complex 
structure on the moduli space induced by the one on $\Sigma$. 
Moreover, we conclude from 
(\ref{omega}) that it is proportional to the K\"ahler form $\omega_{\rm
Sam}$ of Samols' metric \cite{Sam},
\begin{equation} \label{relwwSam}
\omega=-\frac{\kappa}{2}\,\omega_{\rm Sam}.
\end{equation}
It is a global $(1,1)$-form on ${\cal M}_{N}$ with respect to the
complex structure defined by our coordinates.
It becomes apparent that $\omega_{\rm Sam}$ (or $\omega$)
is a central object in the reduced dynamics of both the
aqbelian Higgs model and Manton's model; but we should emphasise
that it plays completely different r\^oles in the two contexts.
In the abelian Higgs model, $\omega_{\rm Sam}$ is the $(1,1)$-form
corresponding to a K\"ahler metric on the configuration space 
${\cal M}_{N}$. In the hamiltonian picture, the dynamics  
takes place on the cotangent bundle $T^{*}{\cal M}_{N}$ with its 
canonical (tautological) symplectic form, and time evolution is
determined by the laplacian of Samols' metric, possibly with an
extra potential term if we allow $\lambda\ne 1$.
On the other hand, in Manton's system $\omega$ is the
symplectic form of a hamiltonian system on ${\cal M}_{N}$
itself, which has no time evolution unless the potential $V^{\rm red}$
is switched on. We can also interpret $\omega$ as the curvature of the
1-form $\cal A$ in (\ref{unitconn}), which represents a unitary
connection on a hermitian line bundle over ${\cal M}_{N}$ in a local
orthonormal frame. This point of view leads directly to the geometric
quantisation of Manton's system using the natural K\"ahler
polarisation, as we shall see in section~6.

\section {Symmetries and conserved quantities}

In this section, we analyse the symmetries of Manton's model on the 
sphere, following the similar analysis in \cite{MNcl} for the plane. More 
precisely, we will study the isometries of the 
metric~(\ref{metric}).
We shall make standard use of Noether's theorem to obtain the 
corresponding 
conserved quantities in the lagrangian formulations of both the field 
theory and the
reduced mechanical system. The conserved quantities reduced to static
solutions are interesting observables of the effective classical system of
vortices, and later we will be concerned with their quantisation.

\subsection {Symmetries in the field theory}

Here, we shall be concerned with the lagrangian (\ref{lagr}).
When computing the Lie derivatives of the different terms along a
vector field $\xi$, one obtains gauge-dependent quantities in
general. However, following \cite{MNcl},
we can supplement the 
field variations under the flow of $\xi$ by a gauge transformation by
$e^{-i\alpha A(\xi)}$, where $\alpha$ is the flow parameter, so as 
to obtain gauge-invariant variations.
The whole operation can be interpreted as a covariant Lie derivative,
in the spirit of the discussion in the Appendix of reference \cite{Sturr}.

The simplest symmetry of the model is time translation, generated by
$\partial_{t}$.  The $O(\alpha)$ variation of the lagrangian is
\[
\alpha \delta {\cal L}=\alpha\left[
\partial_{t}({\cal L}+\gamma A_{t}\Omega^{2}- \mu B A_{t})
+2\mu i\left(\partial_{\bar{z}}(A_{t}E_{z})-\partial_{z}(A_{t}E_{\bar{z}})\right)\right],
\]
where we included a gauge transformation by $e^{-i\alpha A_{t}}$ in
the fields. Noether's theorem then gives the conserved density
\begin{eqnarray*}
j^{t}&=&\sum_{\psi}\frac{\delta {\cal L}}{\delta\partial_{t}\psi}\delta\psi
-\partial_{t}({\cal L}+\gamma A_{t}\Omega^{2}- \mu B A_{t})\\
&=& \frac{1}{2}B^{2}\Omega^{-2}+
(|D_{z}\phi|^{2}+ |D_{\bar{z}}\phi|^{2})
+\frac{\lambda}{8}(1-|\phi|^{2})^{2}\Omega^{2}.
\end{eqnarray*}
This is the density of potential energy, so we learn that $V$ is a
constant of motion. This result does not depend on the particular form
(\ref{conffactor}) for the conformal factor of the metric, and so it
is also valid for more general $\Sigma$.

The $SO(3)$ action on the sphere $\Sigma$ by rotations about axes
through the origin of $\mathbb{R}^{3}$ provides conservation
laws for angular momentum. In our coordinate $z$, this action is
described by elliptic M\"obius transformations with antipodal fixed points,
\begin{equation}\label{Moeb}
z\mapsto \frac{(e^{i\alpha}+|a|^{2})z+a(1-e^{i\alpha})}
{\bar{a}(1-e^{i\alpha})z+(1+|a|^{2}e^{i\alpha})},
\end{equation}
where $a \in \mathbb{C}$ and $\alpha \in \mathbb{R}$. We shall
consider the effect of rotations $R^{(j)}_{\alpha}$ ($j=1,2,3$) 
by an angle $\alpha$ about the three cartesian axes, which 
correspond to taking $a=1,i,0$.

The variation of the lagrangian under the rotation $R^{(1)}_{\alpha}$, to be
supplemented by the gauge transformation $e^{-\frac{\alpha}{2}\left(
(1-z^{2})A_{z}-(1-\bar{z}^{2})A_{\bar{z}}\right)}$, is given up to 
$O(\alpha)$ by
\begin{eqnarray*}
\alpha \delta_{(1)} {\cal L}&=&
\frac{\alpha}{2}
i\left[
\partial_{t} \left[ (\mu B -\gamma \Omega^{2}) 
\left( (1-z^{2}) A_{z}- (1-\bar{z}^{2}) A_{\bar{z}} \right) \right] \right.\\
&&-\left.\partial_{z}\left[(1-z^{2}){\cal L}+2\mu i \left(
(1-z^{2})A_{z}E_{\bar{z}}-(1-\bar{z}^{2})A_{\bar{z}}E_{\bar{z}}\right)
\right] \right. \\
&&\left. +\partial_{\bar{z}}\left[(1-z^{2}){\cal L}+2\mu i\left(
(1-\bar{z}^{2})A_{\bar{z}}E_{z}-(1-z^{2})A_{z}E_{z}\right)
\right] \right],
\end{eqnarray*}
and the variations $\delta_{(2)}{\cal L}$, $\delta_{(3)}{\cal L}$ are
given by similar expressions. The densities of the conserved
quantities can then be shown to be
\begin{eqnarray*}
j^{t}_{(1)}&=&\frac{\gamma i}{2}\left( 
(1-z^{2})(J_{z}+A_{z})-
(1-\bar{z}^{2})(J_{\bar{z}}+A_{\bar{z}})\right) \Omega^{2}\\
j^{t}_{(2)}&=&-\frac{\gamma}{2}\left( 
(1+z^{2})(J_{z}+A_{z})+
(1+\bar{z}^{2})(J_{\bar{z}}+A_{\bar{z}})\right) \Omega^{2}\\
j^{t}_{(3)}&=&-\gamma i\left( z(J_{z}+A_{z})-
\bar{z}(J_{\bar{z}}+A_{\bar{z}})\right) \Omega^{2}.
\end{eqnarray*}
where $J_{z}$ is the supercurrent defined in (\ref{superc}).
These densities are still not gauge invariant. As in \cite{MNcl}, we can
remedy this by adding to the vector field $X_{(k)}$ in 
$\delta_{(k)}{\cal  L}=:\partial_{\nu}X^{\nu}_{(k)}$ a divergenceless
vector field. So we substitute $X^{\nu}_{(k)}$ by 
$\tilde{X}^{\nu}_{(k)}$ given by
\begin{eqnarray*}
\tilde{X}_{(k)}^{t}&=&X_{(k)}^{t}
-\partial_{z}(\Lambda_{(k)}A_{\bar{z}})
+\partial_{\bar{z}}(\Lambda_{(k)}A_{z})\\
\tilde{X}_{(k)}^{z}&=&X_{(k)}^{z}
+\partial_{t}(\Lambda_{(k)}A_{\bar{z}})
-\partial_{\bar{z}}(\Lambda_{(k)}A_{t})\\
\tilde{X}_{(k)}^{\bar{z}}&=&X_{(k)}^{\bar{z}}
-\partial_{t}(\Lambda_{(k)}A_{z})
+\partial_{z}(\Lambda_{(k)}A_{t}),
\end{eqnarray*}
using
\[
\left(\Lambda_{(1)},\Lambda_{(2)},\Lambda_{(3)}\right)=
-2i\gamma R^{2} \left(   
\frac{z+\bar{z}}{1+|z|^{2}},
-i\frac{z-\bar{z}}{1+|z|^{2}},
\frac{1-|z|^{2}}{1+|z|^{2}}
\right).
\]
The new densities $\tilde{j}_{(k)}^{t}$ are now gauge invariant, and
their space integrals are the conserved quantities
\begin{eqnarray*}
M_{1}&=& -\frac{\gamma}{4}\int_{\mathbb{C}}\left[\left( 
(1-z^{2})J_{z}-(1-\bar{z}^{2})J_{\bar{z}}\right)\Omega^{2}
+2iR^{2}\frac{z+\bar{z}}{1+|z|^{2}}B
\right] dz\wedge d\bar{z} \\
M_{2}&=&\frac{\gamma i}{4}\int_{\mathbb{C}}\left[\left(
(1+z^{2})J_{z}+ (1+\bar{z}^{2})J_{\bar{z}}\right)\Omega^{2}
-2iR^{2}\frac{z-\bar{z}}{1+|z|^{2}}B\right] 
dz\wedge d\bar{z}\\
M_{3}&=&\frac{\gamma i}{2}\int_{\mathbb{C}}\left[ i\left(zJ_{z}-
\bar{z}J_{\bar{z}}\right)\Omega^{2}+
2R^{2}\frac{1-|z|^{2}}{1+|z|^{2}}B \right]
dz\wedge d\bar{z}
\end{eqnarray*}
which can be interpreted as angular momenta around the three
independent axes.
Although Noether's theorem determines the conserved 
quantities corresponding to a given symmetry generator only up to an 
additive constant, the requirement that $M_{k}$ be the components of a
hamiltonian moment of $SO(3)$ removes this ambiguity.

Just as in \cite{MNcl}, the quantities $M_{k}$ can be neatly written as
moments of the vorticity of the system. This is defined to be the 
gauge-invariant real quantity
\[
{\cal V}=2i(\partial_{\bar{z}}J_{z}-\partial_{z}J_{\bar{z}})+B,
\]
which is well-defined and smooth everywhere on $\Sigma$. Typically, it
approaches zero away from the vortex cores, where both the magnetic 
field and the supercurrent (and its derivatives) become negligible.
We obtain
\begin{eqnarray}
M_{1}&=&\frac{i\gamma}{2} R^{2}\int_{\mathbb{C}}
\frac{z+\bar{z}}{1+|z|^{2}} \,{\cal V}\,dz\wedge d\bar{z} \label{M1vort}\\
M_{2}&=&\frac{i\gamma}{2} R^{2}\int_{\mathbb{C}}
\frac{(-i)(z-\bar{z})}{1+|z|^{2}} \,{\cal V}\,dz\wedge d\bar{z} \label{M2vort}\\
M_{3}&=&\frac{i\gamma}{2} R^{2}\int_{\mathbb{C}}
\frac{1-|z|^{2}}{1+|z|^{2}}\, {\cal V}\,dz\wedge d\bar{z} \label{M3vort}
\end{eqnarray}
The expressions in the integrands should be compared with the cartesian
coordinates on the sphere as given by equation (\ref{cartes}). We can
anticipate that $\mathbf{M}=(M_{1}, M_{2}, M_{3})$ is a vector in 
$\mathbb{R}^{3}$ which contains information about a centre of mass 
of the vortex configurations.
In particular, when $N$ vortices become coincident, we expect $\mathbf{M}$ 
to point in the direction of the core, given the circular symmetry of
the fields.

\subsection {Symmetries in the effective mechanical system}

There is an action of $SO(3)$ on $\mathbb{CP}^{N}$ by simultaneous 
rotation of the vortex positions $z_{r}$ as in (\ref{Moeb}). It is
generated by the vector fields
\begin{eqnarray}
\xi_{(1)}&=&-\frac{i}{2}\sum_{r=1}^{N}\left((1-z_{r}^{2})
\frac{\partial}{\partial z_{r}}- (1-\bar{z}_{r}^{2})\frac{\partial}
{\partial \bar{z}_{r}}\right) \label{csi1},\\ 
\xi_{(2)}&=&\frac{1}{2}\sum_{r=1}^{N}\left((1+z_{r}^{2})\frac{\partial}
{\partial z_{r}}+
(1+\bar{z}_{r}^{2})\frac{\partial}{\partial \bar{z}_{r}}\right)\label{csi2},\\
\xi_{(3)}&=&i\sum_{r=1}^{N}\left(z_{r}\frac{\partial}{\partial z_{r}}-
\bar{z}_{r}\frac{\partial}{\partial \bar{z}_{r}}\right)\label{csi3},
\end{eqnarray}
which can be seen to extend smoothly to $\mathbb{CP}^{N}$ after changing 
coordinates from the $z_{r}$ to the $s_{k}$ defined in (\ref{esses}). 
The rotational symmetry yields three independent relations among the
functions $b_{r}$, which we shall now derive.
Notice that the fluxes of the vector fields $\xi_{(j)}$ are given by
acting on each $z_{r}$ by the rotations $R^{(j)}_{\alpha}$ introduced 
in section~5.1.
The Lie derivatives of the $b_{r}$ can be computed by making use 
of (\ref{btransf}),
\[
{\rm \pounds}_{\xi_{(j)}}b_{r}=
\lim_{\alpha\rightarrow 0}\frac{R_{\alpha}^{(j)*}b_{r}-b_{r}}{\alpha}.
\]
We then find the following relations:
\begin{eqnarray}
&\D -\frac{i}{2}\sum_{s=1}^{N}\left((1-z_{s}^{2})\frac{\partial b_{r}}{\partial z_{s}}-
(1-\bar{z}_{s}^{2})\frac{\partial b_{r}} {\partial \bar{z}_{r}}\right)=
{\rm \pounds}_{\xi_{(1)}}b_{r}=-i(1+z_{r}b_{r})& \label{Liexi1}\\
& \D \frac{1}{2}\sum_{s=1}^{N}\left((1+z_{s}^{2})\frac{\partial b_{r}} {\partial z_{s}}+
(1+\bar{z}_{s}^{2})\frac{\partial b_{r}}{\partial \bar{z}_{s}}\right)=
{\rm \pounds}_{\xi_{(2)}}b_{r}=-(1+z_{r}b_{r}) &\label{Liexi2}\\
&\D i\sum_{s=1}^{N}\left(z_{s}\frac{\partial b_{r}}{\partial z_{s}}-
\bar{z}_{s}\frac{\partial b_{r}}{\partial \bar{z}_{s}}\right)=
{\rm \pounds}_{\xi_{(3)}}b_{r}=-ib_{r} &\label{Liexi3}
\end{eqnarray}
Using (\ref{bpotential}), equations (\ref{Liexi1}) and (\ref{Liexi2})
can be written as
\[
\sum_{s=1}^{N}\left((1\mp{z}_{s}^{2})\frac{\partial {b}_{s}}
{\partial {z}_{r}} 
\mp(1\mp \bar{z}_{s}^{2})\frac{\partial \bar{b}_{s}}{\partial {z}_{r}}\right)
\mp 2(1+{z}_{r}{b}_{r})=0
\]
which together imply that the quantity
$\sum_{s=1}^{N}(2z_{s}+z_{s}^{2}b_{s}+\bar{b}_{s})$
is constant for all vortex configurations. 
To find what this constant is, we remark that all the singular parts
in (\ref{btil}) cancel in pairs in the sum over $s$, and that
in the limit of coincidence of the vortices the functions 
$\tilde{b}_{r}$ in (\ref{btil}) tend to $b$ in (\ref{bN}).
In particular, when all the vortices are at $Z=0$,
\begin{equation} \label{sumbr}
\sum_{r=1}^{N} b_{r}(0,\ldots,0)=0.
\end{equation}
So we conclude that
\begin{equation} \label{nice12}
\sum_{s=1}^{N}(2z_{s}+z_{s}^{2}b_{s}+\bar{b}_{s})=0.
\end{equation}
Similarly, equation (\ref{Liexi3}) and its conjugate
imply that the quantity 
$\sum_{r=1}^{N}(z_{s}b_{s}-\bar{z}_{s}\bar{b}_{s})$
is independent of the vortex positions. 
Using the explicit formula (\ref{bN}) for $N$ coincident vortices,
we then deduce that this constant has to be zero, obtaining
\begin{equation}\label{nice3}
\sum_{s=1}^{N} z_{s}b_{s} \in \mathbb{R}.
\end{equation}
We remark that equations (\ref{nice12}) and (\ref{nice3}) are
analogues of the statement that the sum of the $b_{r}$ vanishes
for any vortex configuration on the plane, as found by
Samols \cite{Sam} as a consequence of translational symmetry.
In fact, this statement follows from (\ref{nice12}) in the limit where
all vortices approach the origin, and (\ref{sumbr}) may be
regarded as a special case of it. Equation (\ref{nice3}) is also
valid for vortices on the plane, and is a consequence of the $SO(2)$ 
symmetry, but it has not been noted before in the literature.

The $SO(3)$ action on $\mathbb{CP}^{N}$ leaves the symplectic
form (\ref{omega}) invariant, i.e.
\begin{equation}\label{sympaction}
{\rm \pounds}_{\xi}\omega=d\iota_{\xi}\omega=0
\end{equation}
for all $\xi$ generating a rotation. To establish this, we make use
of the relations (\ref{nice12}) and (\ref{nice3}). For example,
$\xi_{(3)}$ satisfies (\ref{sympaction}) if and only if
\[
\frac{\partial}{\partial\bar{z}_{q}}\left[
\sum_{s=1}^{N}\left(\bar{z}_{s}\frac{\partial \bar{b}_{s}}{\partial
z_{r}} - {z}_{s}\frac{\partial {b}_{s}}{\partial z_{r}} \right)
-b_{r}\right]=0
\]
for all $q$ and $r$, and this follows from (\ref{nice3}) or the 
weaker statement (\ref{Liexi3}). Since 
$H^{1}(\mathbb{CP}^{N};\mathbb{R})$ is trivial, (\ref{sympaction}) 
implies that there exist globally defined functions $M_{j}^{\rm red}$ 
satisfying
\begin{equation} \label{Mjred}
\iota_{\xi_{(j)}}\omega=-dM^{\rm red}_{j}.
\end{equation}
The functions $M^{\rm red}_{j}$ are determined from (\ref{Mjred})
only up to a constant, and the $\xi_{(j)}$ are their corresponding
hamiltonian vector fields. We can fix this constant by requiring that
the $M^{\rm red}_{j}$ are the components of a moment map, i.e.~that the
$SO(3)$ action is hamiltonian (cf.~\cite{GS}),
\begin{equation} \label{hamaction}
\{M^{\rm red}_{i},M^{\rm red}_{j}\}:=\omega(\xi_{(i)},\xi_{(j)})=
-\sum_{k=1}^{3}\epsilon_{ijk}M^{\rm red}_{k}.
\end{equation}

The $M_{j}^{\rm red}$ turn out to be the conserved quantities
corresponding to the rotational symmetry in the reduced mechanical
system. Recall that the connection 1-form $\cal A$
in (\ref{unitconn}) is a symplectic potential for $\omega$, and thus
equation (\ref{Mjred}) is equivalent to 
\begin{equation}\label{invconn}
{\rm \pounds}_{\xi_{(j)}}{\cal A}=dW_{j}
\end{equation}
with
\begin{equation} \label{WandM}
W_{j}={\cal A}(\xi_{(j)})-M^{\rm red}_{j}.
\end{equation}
Equation (\ref{invconn}) is the statement of rotational invariance of
the $U(1)$ connection represented by ${\cal A}$
(see~\cite{JM} for a discussion
in the more general situation of connections on bundles with
nonabelian structure group). The reduced lagrangian $L^{\rm red}$ 
has a kinetic term (\ref{redT}) of the form
\[
T^{\rm red}=\sum_{r=1}^{N}{\cal A}_{r}\dot{z}_{r}+ {\rm c.c.}
\]
and a rotationally-invariant potential. Using (\ref{invconn}), we can
establish that
\begin{equation}\label{dWdt}
{\rm \pounds}_{\xi_{(j)}}L^{\rm red}=\partial_{t}W_{j}
\end{equation}
and so Noether's theorem implies that $M_{j}^{\rm red}$ as given by 
(\ref{WandM}) is a conserved quantity. Notice that
$W_{j}$ depends on the choice of symplectic potential for $\omega$, 
whereas $M_{j}^{\rm red}$ does not --- cf. equations (\ref{invconn}) 
and (\ref{Mjred}).
We can determine $M^{\rm red}_{j}$ 
by integrating (\ref{Mjred}), or alternatively from (\ref{dWdt}) 
and (\ref{WandM}). 
We shall follow the latter route, which provides a direct proof
of the spherical symmetry of the reduced system. Using equations
(\ref{Liexi1})--(\ref{Liexi3}),
we find that 
\begin{eqnarray*}
{\rm \pounds}_{\xi_{(1)}}L^{\rm red}&=& 
-\pi \gamma  (R^{2}-N)\partial_{t}\sum_{r=1}^{N}(z_{r}+\bar{z}_{r})\\
{\rm \pounds}_{\xi_{(2)}}L^{\rm red}&=&
\pi i \gamma
(R^{2}-N)\partial_{t}\sum_{r=1}^{N}(z_{r}-\bar{z}_{r})\\
{\rm \pounds}_{\xi_{(3)}}L^{\rm red}&=& 0 \;=\;2 \pi \gamma 
\partial_{t}\left( N(N-R^{2}) \right).
\end{eqnarray*}
The constant term after the time derivative in the last equation was
chosen so that the conserved quantities $M^{\rm red}_{j}$ obey
(\ref{hamaction}). Making use of the relations (\ref{nice12}) and
(\ref{nice3}) for the functions $b_{r}$, they
can be written as
\begin{eqnarray} 
M_{1}^{\rm red}&=&\frac{\kappa}{4}\sum_{r=1}^{N}\left(
2R^{2}\frac{z_{r}+\bar{z}_{r}}{1+|z_{r}|^{2}}+b_{r}+\bar{b}_{r}
\right), \label{simpM1red}\\
M_{2}^{\rm red}&=&\frac{i \kappa}{4}\sum_{r=1}^{N}\left(
-2R^{2}\frac{z_{r}-\bar{z}_{r}}{1+|z_{r}|^{2}}+b_{r}-\bar{b}_{r}
\right), \label{simpM2red}\\
M_{3}^{\rm red}&=&\frac{\kappa}{2}\sum_{r=1}^{N}\left(
R^{2}\frac{1-|z_{r}|^{2}}{1+|z_{r}|^{2}}-(z_{r}b_{r}+1)
\right). \label{simpM3red}
\end{eqnarray}

A consistency check of the reduction procedure can be made by
comparing the conserved angular momenta in the two pictures. To do
this, we shall write the quantities $M_{k}$ in section~5.1 for static 
solutions in terms of the moduli. 
This is most easily done from equations
(\ref{M1vort})--(\ref{M3vort}), expressing the fields in terms of the
function $h$ and making use of equations 
(\ref{newBogom}) and (\ref{bs}) to reduce each
$M_{k}$ to the moduli space, similarly to what we did for the
lagrangian in section~4. 
For example, to obtain the expression for $M_{1}$ we start by writing 
\begin{eqnarray*}
\frac{z+\bar{z}}{1+|z|^{2}}{\cal V}&=&
\frac{2}{R^{2}} \frac{z+\bar{z}}{1+|z|^{2}} \partial_{z}\left(
\frac{\partial_{\bar{z}}h \,\partial_{z}\partial_{\bar{z}}h }
{1+|z|^{2}}\right)\\
&=&\frac{4}{R^{2}}\partial_{z}\left( (z+\bar{z})
\partial_{z}\left(\frac{(\partial_{\bar{z}}h)^{2}}{1+|z|^{2}}\right)
-\frac{(\partial_{\bar{z}}h)^{2}}{1+|z|^{2}} \right) + {\rm c.c.}
\end{eqnarray*}
and use Stokes' theorem to evaluate (\ref{M1vort}) as a sum of contour
integrals along small discs $C_{r}$ of radius $\epsilon$ around the vortex
positions:
\[
M_{1}=\frac{\gamma i}{4}\sum_{r=1}^{N}\oint_{C_{r}}\left(
(z+\bar{z})\left( \bar{z} (\partial_{\bar{z}}h)^{2}+
(1+|z|^{2})\partial_{\bar{z}}h \partial_{z}\partial_{\bar{z}}h  \right)
-(1+|z|^{2})(\partial_{\bar{z}}h)^{2}  \right)d\bar{z}+{\rm c.c.}
\]
Assuming that the vortices are isolated, we may write on $C_{r}$
\begin{eqnarray*}
\partial_{z}\partial_{\bar{z}}h&=&-\frac{R^{2}}{(1+|z_{r}|^{2})^{2}}+
O(\epsilon^{2})\\
\partial_{\bar{z}}h&=&\frac{e^{i\theta_{r}}}{\epsilon}+\frac{\bar{b}_{r}}{2}
+O(\epsilon)
\end{eqnarray*}
and then obtain in the limit $\epsilon\rightarrow 0$
\begin{eqnarray*}
M_{1}&=&\frac{\pi \gamma}{2}\sum_{r=1}^{N} \left(
4R^{2}\frac{z_{r} + \bar{z}_{r}}{1+|z_{r}|^{2}}+
\left( 1-z_{r}^{2}\right)b_{r}+ \left( 1-\bar{z}_{r}^{2}\right)\bar{b}_{r} 
-2(z_{r}+\bar{z}_{r}) 
\right) \\
&=&\frac{\kappa}{4}\sum_{r=1}^{N}\left(
2R^{2}\frac{z_{r}+\bar{z}_{r}}{1+|z_{r}|^{2}}+b_{r}+\bar{b}_{r}
\right),
\end{eqnarray*}
where we made use of (\ref{nice12}).
Similarly, we find
\begin{eqnarray*}
M_{2}&=&\frac{\pi \gamma i}{2}\sum_{r=1}^{N} \left(
-4R^{2}\frac{z_{r} - \bar{z}_{r}}{1+|z_{r}|^{2}}+\left( 1+z_{r}^{2}\right)b_{r}- \left(1+\bar{z}_{r}^{2}\right)\bar{b}_{r}
+2(z_{r}-\bar{z}_{r}) \right)\\
&=&\frac{\kappa i}{4}\sum_{r=1}^{N}\left(
-2R^{2}\frac{z_{r}-\bar{z}_{r}}{1+|z_{r}|^{2}}+b_{r}-\bar{b}_{r}
\right)
\end{eqnarray*}
and
\begin{eqnarray*}
M_{3}&=&{\pi \gamma}\sum_{r=1}^{N} \left(
2R^{2}\frac{1-|{z}_{r}|^{2}}{1+|z_{r}|^{2}}
-\left( z_{r}b_{r}+ \bar{z}_{r}\bar{b}_{r}\right) 
-2 \right)\\
&=&\frac{\kappa}{2}\sum_{r=1}^{N}\left(
R^{2}\frac{1-|z_{r}|^{2}}{1+|z_{r}|^{2}}-(z_{r}b_{r}+1)
\right). 
\end{eqnarray*}
So each $M_{j}$ agrees with $M_{j}^{\rm red}$.

It is instructive to compare the conserved quantities that we have found
for the sphere with the ones obtained for the plane in \cite{MNcl}.
On the plane, space isometries are described by the euclidean group
$E(2)$. Convenient generators are the translations
along the $x_{1}$ and $x_{2}$ axes and the rotation about the origin, and
their conserved quantities in the reduced picture were determined to be
\begin{eqnarray}
P_{1}&=&
-{\pi \gamma i}\sum_{r=1}^{N} 
(Z_{r}-\bar{Z}_{r}),\label{P1plane}\\
P_{2}&=&
\-{\pi \gamma}\sum_{r=1}^{N}(Z_{r}+\bar{Z}_{r}),\label{P2plane}\\
M&=&
\-2\pi\gamma\sum_{r=1}^{N}\left(\frac{1}{2}|Z_{r}|^{2}+B_{r}Z_{r}+
\bar{B}_{r}\bar{Z}_{r} +1 \right),\label{Mplane}
\end{eqnarray}
where $Z_{r}$ denote the positions of the vortex cores, $B_{r}$ the 
coefficients in an expansion equivalent to (\ref{bs}), and we removed
the `red' superscripts.
In the limit where the radius $R$ is large and the vortices are close
together, say in a small neighbourhood of the North pole, one should
expect that our $M_{k}$ should be well approximated by quantities 
directly related to the ones in (\ref{P1plane})--(\ref{Mplane}).
Indeed, identifying $2 R z_{r}=Z_{r}$ we obtain from 
(\ref{simpM1red})--(\ref{simpM3red})
\begin{eqnarray}
M_{1}&=&-RP_{2}+\pi\gamma R\sum_{r=1}^{N}(B_{r}+\bar{B}_{r})+O(|z_{s}|)  
\label {M1andP2} \\
M_{2}&=&RP_{1}+\pi\gamma iR\sum_{r=1}^{N}(B_{r}-\bar{B}_{r})+O(|z_{s}|)  
\label{M2andP1}\\
M_{3}&=&2\pi\gamma NR^{2}-M+O(|z_{s}|). \label{M3andM}
\end{eqnarray}
Since it is known from \cite{Sam} that $\sum_{r=1}^{N}B_{r}=0$
(cf.~equation~(\ref{sumbr})), we see
that $M_{1}$, $M_{2}$ and $P_{1}$, $P_{2}$ are related as 
expected, but the 
equation (\ref{M3andM}) for $M_{3}$ is rather surprising. It means that, as 
$R\rightarrow\infty$, 
$M_{3}$ becomes infinite, and we should subtract from it
the quantity $2\pi\gamma N R^{2}$, itself infinite in the limit, to be
able to compare it with the angular momentum in the plane, $M$.
Notice that when (\ref{M1andP2})--(\ref{M3andM}) are inserted in 
(\ref{hamaction}), we obtain
\[
\{P_{1},P_{2}\}=2\pi\gamma N-\frac{1}{R^{2}}M+O(|z_{s}|).
\]
Thus when $R\rightarrow \infty$ we recover the nonvanishing
classical Poisson brackets for the linear momenta in the plane
as calculated in \cite{Hor}.

For $N$ coincident vortices, we can make use of (\ref{bN}) to obtain the
angular momentum vector in closed form:
\begin{equation}\label{coincidence}
\mathbf{M}^{[N]}=2\pi\gamma N(R^{2}-N)
\left(   
\frac{Z+\bar{Z}}{1+|Z|^{2}},
-i\frac{Z-\bar{Z}}{1+|Z|^{2}},
\frac{1-|Z|^{2}}{1+|Z|^{2}}
\right).
\end{equation}
As before, $Z$ denotes the position of the common core.
If we take $N=1$, (\ref{coincidence}) implies that
a single vortex on $\Sigma$ should be assigned a nonzero $\mathbf{M}$ 
vector. Its direction gives just the vortex position, and its conservation 
implies that the single vortex does not move (even if $\lambda\ne 1$), 
as expected. The length of $\mathbf{M}^{[1]}$ can be interpreted as a
nonzero intrinsic angular momentum of the single vortex at rest. This
was also a feature of the model in the plane, where a single vortex
was found to have an
intrinsic momentum $-2\pi\gamma$. This agrees with our result, provided
that we subtract the ``tail'' momentum $2\pi\gamma N R^{2}$ as discussed
above. It should be noted that on the sphere this intrinsic momentum
is quantised, from the considerations in section~2 --- it is a half 
integer in units of $\hbar=1$. On the
plane, this feature is not apparent, since $\gamma$ could take any
value. More generally, a configuration of $N$ coincident vortices has
angular momentum $-2\pi\gamma N^{2}$ after subtraction of 
$2\pi\gamma N R^{2}$, and this is consistent with the results in \cite{MNcl}.

\section{Ingredients for geometric quantisation}

We would like to investigate the quantum version of the reduced
mechanics in the framework of geometric quantisation.
We shall follow the conventions in 
\cite{Woo} and refer to \cite{GH}
for background on complex geometry. 
To construct the quantum system, we need to supplement the classical 
theory (specified by the phase space 
${\cal M}_{N}=\mathbb{CP}^{N}$, endowed with the symplectic form 
$\omega$) with a hermitian line bundle $L$ over ${\cal M}_{N}$.
The wavefunctions in the quantum Hilbert space are particular 
sections of $L$.

To start with, we should verify whether our phase space is quantisable
at all. This is equivalent to the integrality of the class represented
by the closed form $\frac{1}{2\pi}\omega$ in de Rham cohomology,
\begin{equation} \label{Weil}
\frac{1}{2\pi}[\omega] \in H^{2}({\cal M}_{N};\mathbb{Z})\subset 
H^{2}({\cal M}_{N};\mathbb{R}).
\end{equation}
In general, this requirement leads to nontrivial constraints on the
parameters of the classical theory --- the Weil (pre)quantisation
conditions. If they are satisfied, we may regard
$\frac{1}{2\pi}[\omega]$ as the
first Chern class of a smooth complex line bundle over ${\cal M}_{N}$,
which is what we call the prequantum line bundle $L$.

Recall that $H^{2}(\mathbb{CP}^{N};\mathbb{R})$ is cyclic and we can
take as generator the first Chern class $\eta \in 
H^{2}(\mathbb{CP}^{N};\mathbb{Z})$ of the hyperplane bundle of 
$\mathbb{CP}^{N}$. Then $[\omega] = 2 \pi \ell \eta$ for suitable $\ell
\in \mathbb{R}$. To determine $\ell$, we can refer to equation
(\ref{relwwSam}) and use the formula for the cohomology class of 
$\omega_{\rm Sam}$ obtained by Manton in \cite{Msmv}. (This formula
has been generalised in \cite{MNvvms} for $\Sigma$ of arbitrary genus.) For
the benefit of the reader, we reproduce Manton's argument here. Let
${\cal M}_{N}^{\rm co}\subset {\cal M}_{N}$ be the subvariety of
configurations of $N$ coincident vortices. This is a projective line
parametrised by the position $Z$ of the zero of the Higgs field.
Equation (\ref{bN}) implies that $\omega$ restricts to it as
\begin{equation}\label{wcoinc}
\omega|_{{\cal M}^{\rm co}_{N}}=-i\kappa \frac{N(R^{2}-N)}
{\left(1+|Z|^{2}\right)^{2}}
dZ\wedge d\bar{Z}.
\end{equation}
It is readily seen that ${\cal M}^{\rm co}_{N}$ is embedded as a projective
curve of degree $N$ in ${\cal M}_{N}$, and is thus homologous to
$\pm N [\mathbb{CP}^{1}]$. 
Here, we denote by $[\mathbb{CP}^{1}]$ the homology class of a
projective line inside ${\cal M}_{N}$, which is dual to $\eta$ and 
a generator of $H_{2}({\cal M}_{N};\mathbb{Z})$.
The integral of (\ref{wcoinc}) over 
${\cal M}^{\rm co}_{N}$ is just $- 2 \pi \kappa N(R^{2}-N)$, and so
we conclude that $\ell=-\kappa(R^{2}-N)$,
\begin{equation} \label{ell}
\frac{1}{2\pi}[\omega]=\ell \eta=-\kappa (R^{2}-N)\eta.
\end{equation}
Equation (\ref{Weil}) is equivalent to $\ell$ being an integer,
\[
\kappa (R^{2}-N)\in \mathbb{Z}
\]
and this is weaker than the conditions 
$\kappa, \kappa R^{2}\in \mathbb{Z}$ 
that we already had to impose in (\ref{mu}) and (\ref{consist}) from 
considerations of gauge invariance (${\rm mod\;}2\pi$) of the classical 
field theory action. We conclude that no further constraints arise 
from prequantisation.

In geometric quantisation, the prequantum line bundle $L$ is to be
equipped with a hermitian metric and a unitary connection. The
fact that $\mathbb{CP}^{N}$ is simply-connected implies that in our
case $L$ is uniquely determined as a smooth bundle by the symplectic
structure, and so is the
hermitian metric and the connection ${\cal D}$. The basic idea in
the standard construction of $L$ is to interpret (real) symplectic 
potentials of $\omega$ as local expressions for the connection, and then use 
parallel transport to define local sections and construct the 
bundle (cf.~\cite{Woo}). 
A given symplectic potential determines a unique local section 
$\sigma$ of $L$ up to a phase of modulus one.
The hermitian metric is introduced by requiring that each $\sigma$ is
a local orthonormal frame,
\begin{equation}\label{hermit}
\langle \sigma,\sigma \rangle =1.
\end{equation}
This is unambiguous since two symplectic potentials must differ by the
exterior derivative of a real function $u$, and then the corresponding 
local sections are related by the factor $e^{-iu}$.

The wavefunctions in geometric quantisation are defined as the
$L^{2}$ polarised sections of $L$. By $L^{2}$ we mean
square-integrable with respect to the hermitian product (\ref{hermit})
on the fibres and the symplectic measure $\frac{\omega^{N}}{N!}$ 
on the base ${\cal M}_{N}$.
Roughly speaking, polarised means that they only depend on half of
the real coordinates of the phase space, just as the wavefunctions in
the Schr\"odinger representation of quantum mechanics only depend on
the position and not on the momentum. More precisely, a polarisation 
${\cal P}$ is defined as a lagrangian (i.e.~maximally isotropic)
integrable subbundle of the complexification $T_{\mathbb{C}}{\cal M}_{N}$
of the tangent bundle of the phase space, 
and the condition
\begin{equation}\label{polarised}
{\cal D}_{\bar{X}}\psi=0,\;\;\;\;\;\;\;
\forall \;X\in \Gamma({\cal M}_{N},{\cal P})
\end{equation}
defines what is meant for a section $\psi$ to be 
${\cal P}$-polarised. 
When the classical dynamics takes place in a K\"ahler phase space, as
is our case, there is a natural choice of polarisation ${\cal P}$ --- 
namely, the
one determined by the $i$-eigenspaces of the compatible complex 
structure. It is generated by the holomorphic vector fields
in the local complex coordinates.
The introduction of a K\"ahler polarisation can be interpreted 
naturally in terms of complex geometry as
follows. A connection on the prequantum line bundle defines a
holomorphic structure for $L$: By definition, the holomorphic sections
are the ones which are annihilated by the part of ${\cal D}$ that
takes values in $\Omega^{(0,1)}({\cal M}_{N},L)$, which is defined
from the complex structure in the base. But such sections are
precisely the ones satisfying the condition (\ref{polarised}) for the
K\"ahler polarisation.
Thus, polarised sections of $L$ are nothing but holomorphic sections
with respect to the holomorphic structure on $L$ induced by the
unitary connection~${\cal D}$.

\section{The quantum Hilbert space}

The Picard variety of $\mathbb{CP}^{N}$ is trivial, and this implies that
$L$ is uniquely determined as a holomorphic line bundle by its first
Chern class, which can be read off from (\ref{ell}).
A classical result on the sheaf cohomology of $\mathbb{CP}^{N}$ 
establishes that $L$ admits nontrivial global
holomorphic sections if and only if $\ell>0$ (i.e. $\kappa<0$), and
then they form the vector space (cf.~\cite{Har})
\begin{equation}\label{sections}
H^{0}(\mathbb{CP}^{N}, {\cal O}(L))
\cong \mathbb{C}[Y_{0},\ldots,Y_{N}]_{\ell}
\end{equation}
where the right-hand side denotes the homogeneous polynomials of 
degree $\ell$ in the $N+1$ variables $Y_{j}$. 
This gives a concrete way to realise $L$ and its sections (up to
multiplication by a constant in $\mathbb{C}^{\times}$). Recall that
the local symplectic potential ${\cal A}$ in (\ref{unitconn}) for the 
connection ${\cal D}$ determines a nonvanishing local section 
$\sigma: W_{0}\rightarrow L$. It is not holomorphic though, as 
${\cal A}$ has a nonzero component in 
$\Omega^{(0,1)}(W_{0})$. But we can obtain a holomorphic local section
from it by using a nonunitary gauge transformation: Since
\begin{equation}\label{sigtosig0}
{\cal A} = 2 \pi  \gamma i \sum_{r=1}^{N}
\left( 2 R^{2}\frac{\bar{z}_{r}}{1+|z_{r}|^{2}}+\tilde{b}_{r}\right)dz_{r}
-2\pi \gamma i d\left(
\frac{1}{2}{\cal B}+R^{2}\sum_{r=1}^{N}{\rm log}\left( 1+|z_{r}|^{2}\right)
\right),
\end{equation}
where ${\cal B}$ is defined up to an additive real constant by 
(\ref{bpotential}), we can define on $W_{0}$
\begin{equation}\label{holsec}
\sigma^{(0)}(z_{1}\ldots,z_{N}):=\sigma(z_{1}\ldots,z_{N})e^{-2\pi\gamma
\left(
\frac{1}{2}{\cal B}+R^{2}\sum_{r=1}^{N}{\rm log}\left( 1+|z_{r}|^{2}\right)
\right)};
\end{equation}
this is a holomorphic section of $L$ on $W_{0}$. It is uniquely
determined from $\sigma$ up to a positive real constant, and thus from
${\cal A}$ up to a constant in $\mathbb{C}^{\times}$. It extends to a global
section of $L$, vanishing in the complement of $W_{0}$;
we identify it with
the homogeneous polynomial $Y_{0}^{\ell}$ in (\ref{sections}). From
it, we can define the sections
\[
\sigma^{(j)}:=\left(\frac{Y_{j}}{Y_{0}}\right)^{\ell}\sigma^{(0)}=
\left( y_{j}^{(0)}  \right)^{\ell}\sigma^{(0)}
\]
which trivialise $L$ on each $W_{j}$, and determine the line bundle
through the transition functions
\begin{eqnarray*}
\varphi_{ij}\;:\;W_{i}\cap W_{j}&\longrightarrow &\mathbb{C}^{\times}\\
(y^{(i)}_{1},\ldots,y^{(i)}_{N})&\longmapsto&
\left(y_{j}^{(i)}\right)^{\ell}=
\left(\frac{s_{j}}{s_{i}}\right)^{-\kappa(R^{2}-N)}.
\end{eqnarray*}
On each $W_{j}$, global holomorphic sections of $L$ are given by multiplying
$\sigma^{(j)}$ by polynomials in the $y^{(j)}_{k}$ of degree
less than or equal to $\ell$.

The quantum Hilbert space ${\cal H}_{\cal P}$ is the space
of holomorphic sections of $L$ which are normalisable with respect to
the inner product defined by the symplectic measure of 
${\cal M}_{N}$ and the product on the fibres given by (\ref{hermit}),
as we said in section~6.
This inner product can be easily written down as an integral over the open
dense $W_{0}$, where $L$ is trivialised by $\sigma^{(0)}$, by 
making use of (\ref{omega}), (\ref{hermit}) and (\ref{holsec}).
Since we are dealing with a compact phase space, all the holomorphic 
sections have finite norm, so the Hilbert space ${\cal H}_{\cal P}$ is 
$H^{0}(\mathbb{CP}^{N}, {\cal O}(L))$ itself, with dimension
\begin{equation} \label{dimH}
\dim {\cal H}_{\cal P} = 
\left(
\begin{array}{c}
{N+\ell} \\
{\ell}
\end{array}
\right).
\end{equation}
All these quantum states belong to a single degenerate energy
level when $\lambda=1$. Recall that in this situation the hamiltonian 
vanishes and no motion occurs at the classical level.

We may interpret the expression (\ref{dimH}) as giving the number of 
states in a
quantum system of $N$ interacting bosons. By interacting, we mean that
the area available for the dynamics on the sphere is affected by
the space which the vortices themselves occupy. Recall that 
Bradlow's bound (\ref{Brad}) establishes that $N$ vortices
can only live on a sphere which has an area exceeding $4\pi
N$. Heuristically, we can say that a single vortex occupies $4\pi$
units of area. So we can regard (\ref{dimH}) as the formula for the
number of states for a system of $N$
bosons which can be assigned to any of the 
$\frac{|\kappa|}{4\pi}( 4\pi R^{2}- 4\pi N)$ states
corresponding to the room available on the sphere, after the total area
of the vortices has been discounted. (For $\kappa=-1$, there is a
similiar interpretation for (\ref{dimH}) as the number of states
of a system of $N$ noninteracting fermions, but it breaks down for 
$\kappa\ne -1$.)

From the formula (\ref{ell}), it is immediate to compute the
volume of the moduli space determined by the K\"ahler form $\omega$:
\[
{\rm Vol}_{\omega}({\cal M}_{N})=
\frac{\left(2\pi|\kappa|(R^{2}-N) \right)^{N}}{N!}
=\frac{(2\pi \ell)^{N}}{N!}.
\]
It is of course proportional to the volume determined by Samols'
metric, as first computed by Manton (cf.~\cite{Msmv},\cite{MNvvms}).
This volume has been used to deduce the thermodynamics of an
ideal gas of abelian Higgs vortices at $\lambda=1$ in the framework 
of Gibbs' classical statistical mechanics. In Manton's model at $\lambda=1$, 
there is only a ground state as we noted above, and its degeneracy, in
Gibbs' approximation, is given by
\begin{equation}\label{VolM_N}
d_{\rm Gibbs}=\frac{1}{(2\pi\hbar)^{N}}{\rm Vol}({\cal M}_{N})
=\frac{\ell^{N}}{N!}.
\end{equation}
Notice that Planck's constant is $2\pi \hbar=2\pi$ in our units.
Gibbs' partition function is simply $Z_{\rm Gibbs}=
d_{\rm Gibbs}e^{-\beta N \pi}$.
At $\lambda\ne 1$, the degeneracy is lifted but the formula above is
still to be interpreted as the total number of states of the system.
It is of interest to study the quotient
\[
Q:=\frac{\dim {\cal H}_{\cal P}}
{d_{\rm Gibbs}}
\]
which gives information about how appropriate Gibbs' estimate for the
number of states of the quantum system is. 
From (\ref{dimH}) and (\ref{VolM_N}), we find
\[
Q=\frac{(N+\ell)!}{\ell^{N}\ell!}.
\]
Using Stirling's formula for the gamma function, we obtain
\begin{equation} \label{Qasymp}
Q=\left(1+\frac{N+1}{\ell} \right)^{N}
\left(1+\frac{N}{\ell +1}\right)^{-\frac{1}{2}}
\left[\left(1+\frac{N}{\ell+1}\right)^{\ell+1}e^{-N}\right]
e^{J(N+\ell+1)-J(\ell+1)},
\end{equation}
where $J$ is the asymptotic series
\[
J(z)=\sum_{n=1}^{\infty}\frac{B_{n}}{(2n-1)2n}\frac{1}{z^{2n-1}}.
\]
In the context of Chern--Simons theories, the classical approximation
is described as the limit $|\kappa|\rightarrow \infty$; this 
is equivalent to keeping the coupling $\mu$ as constant and letting
$\hbar\rightarrow 0$. So we keep $N$ fixed and let 
$\ell \rightarrow \infty$ in the expression
(\ref{Qasymp}), and this gives indeed $Q\rightarrow 1$.
We might also try to obtain a classical regime in a thermodynamical
limit, where both $N$ and the area of the sphere become very large,
but keeping a finite (possibly small) density, which we might want
to define as
\[
\nu:=\frac{|\kappa| N}{\ell}=
\frac{\frac{N}{R^{2}}}{\left(1-\frac{N}{R^{2}} \right)}.
\]
But it follows from (\ref{Qasymp}) that in this limit $Q$ is infinite,
however small $\nu$ is taken to be.

\section{Quantum angular momenta}

From the prequantisation data, it is possible to 
construct prequantum operators ${\cal P}(f)$ for any classical observable 
$f\in C^{\infty}({\cal M}_{N})$ as
\begin{equation}\label{preqP}
{\cal P}(f):=-i{\cal D}_{\xi_{f}}+f.
\end{equation}
Here, $\xi_{f}$ is the hamiltonian vector field of $f$ with respect 
to $\omega$, defined by 
\begin{equation}\label{hamvf}
d\iota_{\xi_{f}}\omega=-df.
\end{equation}
Equation (\ref{Mjred}) is of course a special case of (\ref{hamvf}),
with $\xi_{M^{\rm red}_{j}}=\xi_{(j)}$.
In general, the linear operator ${\cal P}(f)$ does not map polarised
sections of $L$ to polarised sections. It is easy to show that it does
if and only if $\xi_{f}$ preserves the polarisation:
\begin{equation}\label{polpreserv}
[\xi_{f},\Gamma({\cal M}_{N},{\cal P})]\subset\Gamma({\cal M}_{N},{\cal P}).
\end{equation}
Then we may interpret ${\cal P}(f)$ as the quantum operator
corresponding to the observable $f$.
In the K\"ahler case, (\ref{polpreserv}) can be seen to be 
equivalent to $\xi_{f}$ being the real part of a holomorphic vector field.
This condition is true for the hamiltonian vector fields 
(\ref{csi1})--(\ref{csi3}) of the
angular momenta in (\ref{simpM1red})--(\ref{simpM3red}).

We can determine explicitly the action of the
quantum operators on the wavefunctions $\Psi$ in the quantum Hilbert
space ${\cal H}_{\cal P}=H^{0}({\cal M}_{N},{\cal O}(L))$. 
In the holomorphic frame on $W_{0}$ provided by $\sigma^{(0)}$, 
one can write $\Psi=\Psi^{(0)}\sigma^{(0)}$ with
\begin{equation}\label{Psi}
\Psi^{(0)}(z_{1},\ldots,z_{N})=\sum
_{j_{1}+\cdots+j_{N}=0}
^{\ell} \alpha_{j_{1}\ldots j_{N}}
\prod_{k=1}^{N}s_{k}^{[N]}(z_{1},\ldots,z_{N})^{j_{k}}
\end{equation}
with $\alpha_{j_{1}\ldots j_{N}} \in \mathbb{C}$. The 1-form representing
${\cal D}$ with respect to this frame can be read off from
(\ref{sigtosig0}) to be
\[
{\cal A}^{(0)} = 2 \pi  \gamma i \sum_{r=1}^{N}
\left( 2 R^{2}\frac{\bar{z}_{r}}{1+|z_{r}|^{2}}+\tilde{b}_{r}\right)dz_{r}.
\]
Substitution in (\ref{preqP}) now gives the local representatives of 
the quantum operators in the local frame $\sigma^{(0)}$.
For example, for $M_{3}$ we obtain
\begin{eqnarray*}
{\cal P}(M_{3})&=&-i\left(\iota_{\xi_{(3)}}d-
i{\cal A}^{(0)}(\xi_{(3)})\right)+M_{3}\\
&=&-i\iota_{\xi_{(3)}}d+\frac{\kappa}{2} N(R^{2}-1)-\kappa
\sum_{r=1}^{N}\sum_{s\ne r}^{N}\frac{z_{r}}{z_{r}-z_{s}}\\
&=&-i\iota_{\xi_{(3)}}d-\frac{N \ell}{2}.
\end{eqnarray*}
Acting on $\Psi$ as in (\ref{Psi}), this yields
\begin{equation}\label{spec}
{\cal P}(M_{3})\Psi^{(0)}=\sum_{j_{1}+\cdots +j_{N}=0}^{\ell}
\left(j_{1}+2j_{2}+\cdots + Nj_{N}-\frac{N\ell}{2} \right)
\alpha_{j_{1}\ldots j_{N}}
\prod_{k=1}^{N}s_{k}^{[N]}(z_{1},\ldots,z_{N})^{j_{k}}.
\end{equation}
From this expression, it is easy to read off the eigenvalues of 
${\cal P}(M_{3})$ as
\[
-\frac{N\ell}{2},-\frac{N\ell}{2}+1,\ldots,\frac{N\ell}{2},
\]
together with their multiplicities. The same spectrum is obtained for
${\cal P}(M_{1})$ and ${\cal P}(M_{2})$.

For $N=1$ and a given negative $\kappa\in \mathbb{Z}$, we see that the
Hilbert space ${\cal H}_{\cal P}$ yields the irreducible (projective)
$(\ell+1)$-dimensional representation of $SO(3)$ through the action of
the generators $M_{j}^{\rm red}$. The situation here is exactly equivalent to
the geometric quantisation of the spin degrees of freedom of a
particle of spin $\frac{\ell}{2}$, which are described classically by
a 2-sphere of half-integer radius $\frac{\ell}{2}$ and the standard 
Fubini--Study symplectic form. More generally, for any $N$, it follows
from (\ref{spec}) that the representation of $SO(3)$ carried by 
${\cal H}_{\cal P}$ is the $N$th symmetric power 
${\rm Sym}^{N}(\boldsymbol{\ell+1})$; notice that $\ell$ itself
depends on $N$. This indicates once again that the vortices in our model
can be regarded as interacting bosons, as we have put forward in 
section~7. It is worthwhile to emphasise how our approach
differs from the usual treatment of a system of indistinguishable
bosons in quantum 
mechanics. In the latter context, the $N$-particle sector of the Fock 
space is constructed as the $N$th symmetric power of the
Hilbert space of a single particle. In our situation, the $N$-particle
sector is constructed directly from the quantisation of a classical
$N$-particle phase space.

\section{Discussion}

In this paper, we have investigated an effective quantisation of
Manton's model of first-order Chern--Simons vortices on a sphere 
$\Sigma$ of 
radius $R$. We have seen that the nontrivial topology of the space
manifold leads to the integer constraints (\ref{mu}) and (\ref{gamma})
on the parameters $\gamma$ and $\mu$ in the lagrangian.
The periodic motion in the classical field theory was reduced
to a hamiltonian system on the moduli space of $N$ vortices
imposing the condition $\gamma=\mu$. At the self-duality point 
$\lambda=1$, the effective dynamics is frozen, whereas for 
$\lambda\simeq 1$ the vortices move
slowly, preserving their energy and angular momenta. The energy is purely 
potential and depends on the relative position of the vortices only.

The angular momenta along the three cartesian axes have been computed 
in section~5, both for the field theory and the reduced dynamics,
and the two results were shown to be consistent. In the latter
context, the expressions for the angular momenta can be simplified
using the relations (\ref{nice12}) and (\ref{nice3}) for the functions
$b_{r}$, which we derived from rotational symmetry. The angular momenta
along the three independent directions fit together to form a moment
map $\mathbf{M}^{\rm red}$ which we can regard as taking values in
an $\mathfrak{so}(3)^{*} \cong \mathbb{R}^{3}$ where the sphere $\Sigma$
is embedded. The direction of the vector $\mathbf{M}^{\rm red}$
gives a point on the sphere that can be interpreted as the
centre of mass of the configuration of $N$ vortices. 
Notice that there is no natural notion of
centroid for a configuration of $N$ points on a sphere (unless
they lie on the same great circle and are not equidistant). We might be
tempted to define it for generic configurations as the direction of the 
sum of the points,
regarded as vectors in an $\mathbb{R}^{3}$ containing the sphere.
(This definition is better behaved if we replace the target
sphere by the elliptic plane by identifying antipodal points.)
For a configuration of vortices, this centroid does not coincide
with the direction of the angular momentum, as can be seen from our
formula for $\mathbf{M}^{\rm red}$ in (\ref{simpM1red})--(\ref{simpM3red}).  
We believe that, for configurations where the vortices are not symmetrically
distributed, the areas of the sphere where vortices are most close
together give a contribution to the angular momentum which is smaller
than the one corresponding to taking the sum of the vortex positions; this is
based on the fact that the total angular momentum of $N$ coincident 
vortices is proportional to $N(R^{2}-N)$ rather than to $N$, as was
shown at the end of section~5.2. On the plane \cite{MNcl}, the
total linear momentum in Manton's model is
proportional to the ordinary centroid in $\mathbb{R}^{2}$ of the 
positions of the vortices.

A rather unexpected feature of the analysis in section~5 is that the
angular momentum of a given number of vortices grows with the square
of the radius $R$ of the sphere. In the limit where the vortices are
kept close together and $R\rightarrow \infty$, the modulus of the
angular momentum blows up, and it was found necessary to subtract the
constant $2\pi\gamma N R^{2}$, which becomes infinite in the limit, in 
order to compare it with the angular momentum of the system of
vortices on the plane. This constant was seen to be related to the
central charge for the linear momentum Poisson algebra for 
Manton's model on the plane.

The geometric quantisation of the reduced model is rather
straightforward to set up. The prequantum line bundle is uniquely 
determined by the K\"ahler structure on the moduli space defined by 
the kinetic energy term. To construct it, we made use of an 
argument of Manton \cite{Msmv} to obtain the cohomology class of  
Samols' K\"ahler 2-form on the moduli space of Bogomol'ny\u\i\
vortices, which appears in the study of the abelian Higgs model. 
It is presumed that the quantum system we have obtained 
approximates a finite truncation of the quantum field theory,
in which most of the excitations are kept in the ground state.
However, it is not clear how one should assess the validity of
this approximation.
For $\lambda=1$, the quantisation of the reduced system yields a
single degenerate energy level; this degeneracy is lifted when the
potential becomes nontrivial, and in principle its spectrum can be 
determined using degenerate perturbation theory.
In section~7, we have computed the dimension of the
quantum Hilbert space and it was shown that it approaches
Gibbs' estimate for the number of quantum states, as determined by the 
volume of the moduli space, in the classical limit of large 
Chern--Simons coefficient. Another result which comes from the
analysis of the quantised effective system is that the solitons in
the model should be interpreted as interacting bosons with the
characteristic size $4\pi$, as explained in section~7. The bosonic
character of the vortices is also apparent from the analysis of
the representations of $SO(3)$ arising in the algebra of the quantum
angular momentum operators. In section~8, we found that 
for $N$ vortices the Hilbert space is the $N$th symmetric power of an
irreducible representation of $SO(3)$. This irreducible representation
is the same as the one obtained from quantising a single vortex on a 
sphere whose area is the one of the original sphere minus the total area 
occupied by $N$ vortices.

\vspace{2cm}
\noindent
{\large \bf{Acknowledgements}}
\vspace{.5cm}

\noindent
The idea of studying the K\"ahler quantisation of Manton's model
of Chern--Simons vortices was first suggested by G.W.~Gibbons.
I am thankful to Nick Manton for many helpful discussions.
The author is supported by Funda\c{c}\~ao para a Ci\^encia e a
Tecnologia, Portugal, through the research grant BD/15939/98, and
thanks Queens' College, Cambridge, for a Munro Studentship.

\end{document}